\documentclass[preprintnumbers,amsmath,amssymb,superscriptaddress]{revtex4}

\usepackage{calc}
\usepackage{graphicx} 
\usepackage{bm}			
\usepackage{color}
\usepackage{amsmath}
\usepackage{amssymb}

\newcommand{\pa}{\partial}
\newcommand{\pd}[2]{\frac{\partial #1}{\partial #2}}
\newcommand{\pdds}[2]{\frac{\partial^2 #1}{\partial #2^2}}

\newcommand{\esp}[1]{\mathbb{E}\left[#1\right] }
\newcommand{\Var}{\mathrm{Var}}

\def\R{{\mathbb R}}

\begin{document}

\title{Exit time distribution in spherically symmetric two-dimensional domains}

\author{J.-F. Rupprecht}
\affiliation{Sorbonne Universit\'es, UPMC Univ Paris 06, UMR 7600, Laboratoire de Physique Th\'eorique de la Mati\`ere Condens\'ee, 4 Place Jussieu, 75005, Paris, France.}

\author{O. B\'enichou}
\affiliation{Sorbonne Universit\'es, UPMC Univ Paris 06, UMR 7600, Laboratoire de Physique Th\'eorique de la Mati\`ere Condens\'ee, 4 Place Jussieu, 75005, Paris, France.}

\author{D. S. Grebenkov}
\affiliation{Laboratoire de Physique de la Mati\`ere Condens\'ee (UMR 7643),
CNRS -- Ecole Polytechnique, F-91128 Palaiseau Cedex France}

\author{R. Voituriez}
\affiliation{Sorbonne Universit\'es, UPMC Univ Paris 06, UMR 8237, Laboratoire Jean Perrin, 4 Place Jussieu, 75005, Paris, France.}
\affiliation{Sorbonne Universit\'es, UPMC Univ Paris 06, UMR 7600, Laboratoire de Physique Th\'eorique de la Mati\`ere Condens\'ee, 4 Place Jussieu, 75005, Paris, France.}


\date{\today}


\begin{abstract} 
The distribution of exit times is computed for a Brownian particle in
spherically symmetric two-dimensional domains (disks, angular sectors,
annuli) and in rectangles that contain an exit on their boundary.  The
governing partial differential equation of Helmholtz type with mixed Dirichlet-Neumann
boundary conditions is solved analytically.  We propose both an exact
solution relying on a matrix inversion, and an approximate explicit
solution.  The approximate solution is shown to be exact for an exit
of vanishing size and to be accurate even for large exits.  For
angular sectors, we also derive exact explicit formulas for the
moments of the exit time.  For annuli and rectangles, the approximate
expression of the mean exit time is shown to be very accurate even for
large exits.  The analysis is also extended to biased diffusion.
Since the Helmholtz equation with mixed boundary conditions is
encountered in microfluidics, heat propagation, quantum billiards, and
acoustics, the developed method can find numerous applications beyond
exit processes.
\end{abstract}

\maketitle

\vskip 1cm
\section{Introduction}

First passage time (FPT) processes are ubiquitous in physics,
chemistry, and biology, with numerous examples of applications ranging from enzymes
searching for specific DNA sequences to animal foraging 
\cite{Redner:2001a,Condamin:2007zl,Benichou:2011,Sheinman2012}. The problem of finding the FPT distribution has direct implications in the fields of neutron or light scattering \cite{Mazzolo2004a} and in
biological modelling.  For instance, the time
needed for an ion to find an open channel is a limiting step in the
kinetics of the neurological process of phototransduction
\cite{Reingruber:2009a}.  The role of the confining domain on the FPT
distribution of a regulation protein to a specific DNA site can
account for the bursting dynamics in gene regulation
\cite{Meyer:2012a}.  

When a target is located on the
boundary of a confining domain, the FPT can be understood as the first
exit time from the domain through an opening (e.g., a ``hole'') on the
boundary.  Hitherto most studies have focused on the mean first
passage time (MFPT) of a Brownian particle to a small exit, which is called the narrow escape problem
\cite{Singer:2006b,ward,al:2011,Isaacson2013}.  For
a starting position which is far enough from the boundary, the FPT
distribution was shown to be dominated by its exponential tail in the limit of a large confining volume: hence
the MFPT was sufficient to characterize the whole FPT distribution,
except for the very short--times region \cite{Benichou:2010a,Meyer:2011,Benichou:2011}.  Note that
the short-time behavior of this distribution was approximately
accounted for by a Dirac distribution whose contribution vanished in
the small exit limit.  A generic multi-exponential representation of
the FPT distribution in domains with heterogeneous distribution of
targets was proposed in \cite{Nguyen2010}.  Some progress to precisely
describe the short-time behavior of the FPT distribution has been
recently achieved.  For instance, Isaacson and Newby proposed a
uniform asymptotic approximation of the FPT distribution in the small
exit limit for 3D confining domains \cite{Isaacson2013}.

In the present article we address the following question: what is the
distribution of the first passage time to an arbitrarily large exit?
To answer this question, we consider a particle diffusing in a
confined spherically symmetric two-dimensional domain $\mathrm{\Omega}
\subset \mathbb{R}^2$ which is periodic along the angular coordinate
$\theta$ and bounded in the radial coordinate $r$.  Examples of such
domains are disks, angular sectors, and annuli.  The analysis is also
applicable to rectangles.  The boundary of $\mathrm{\Omega}$ is
reflecting except for an absorbing patch on the surface through which
the particle can escape.  In spherically symmetric 2D domains, a
Fourier expansion of the survival probability along the periodic
coordinate $\theta$ can be performed.  We adapt the resolution schemes
described in \cite{Sneddon1966} to solve the Helmholtz equation with
mixed boundary conditions satisfied by the survival probability.  Our
approach leads to both exact and approximate expressions for the FPT
distribution and for the moments of the exit time
(Sec. \ref{sec:resolution}).  As a result, we managed to describe the whole distribution of first passage times and
their moments for the escape problem with arbitrary exit size.  The
approximate solution, which is shown to be exact in the limit of a
target of vanishing width, is in fact accurate over the whole range of times scales
even for large exit sizes.  

\begin{figure}[t]
\centering
\includegraphics[width=15cm]{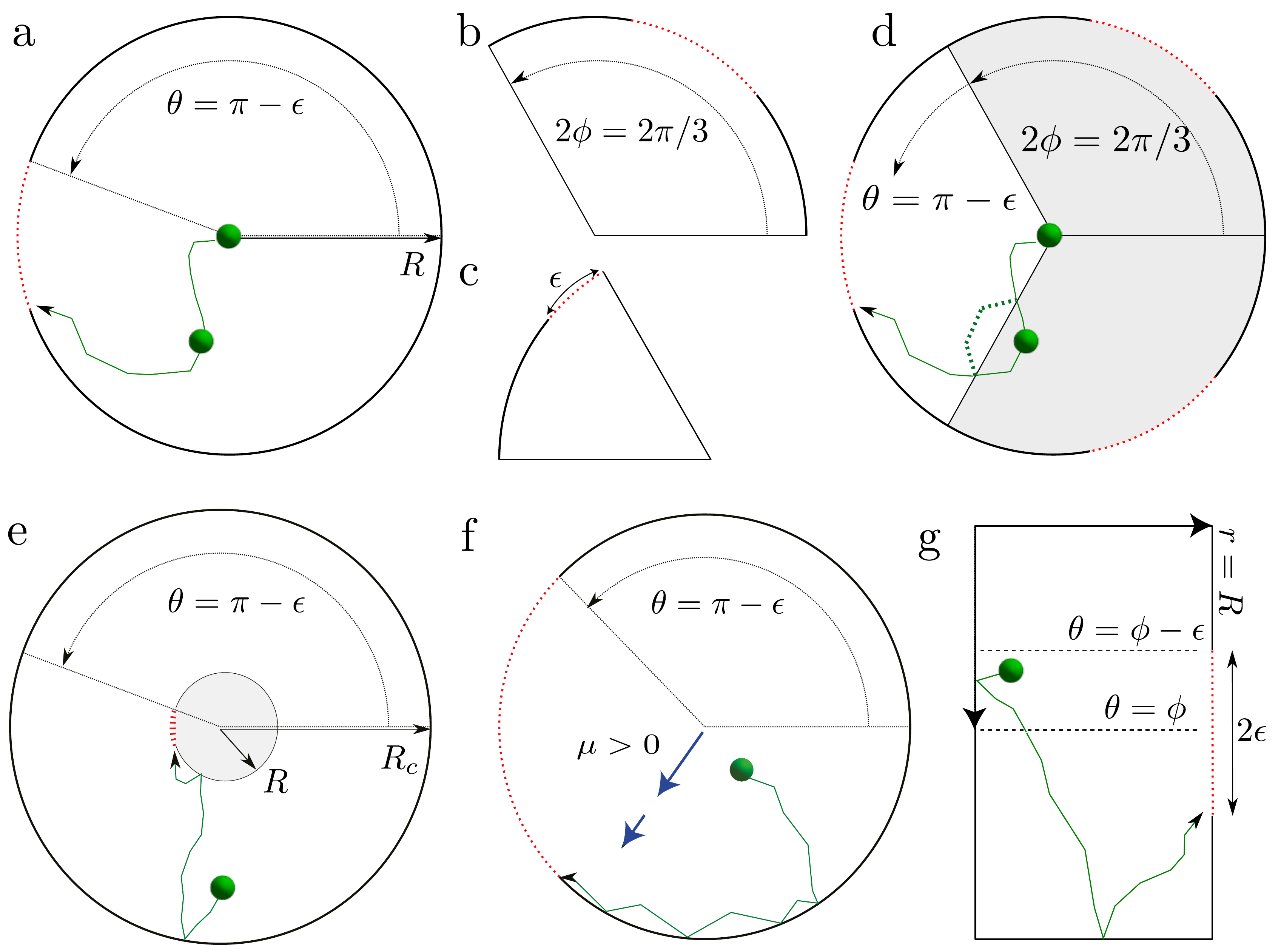}
\caption{
(Color Online) A Brownian particle (green circle) diffuses in a domain
$\mathrm{\Omega}$ whose boundary is reflecting (solid line) except for
an absorbing exit (dashed red line).  \textbf{(a)} $\mathrm{\Omega}$
is a disk of radius $R$ and the exit arclength is $2 R \epsilon$; 
\textbf{(b)} $\mathrm{\Omega}$ is an angular sector of half-aperture
$\phi = \pi/3$, with reflecting rays at $\theta = 0$ and $\theta =
2\phi$; 
\textbf{(c)} $\mathrm{\Omega}$ is an angular sector of total
aperture $\phi = \pi/3$, with reflecting rays at $\theta = 0$ and
$\theta = \phi$ (the exit arclength is $R \epsilon$);
\textbf{(d)} $\mathrm{\Omega}$ is a disk with $3$ regularly spaced
exits of arclength $2 R \epsilon$ on the boundary.  The distribution
of exit times through any of the three regularly spaced exits (case
\textbf{d}) is identical to the distribution of exit times through a
single centered exit within an angular sector of half-aperture $\phi =
\pi/3$ (case \textbf{b}).  We sketch the reflection principle by
representing the trajectory of the particle inside the disk with three
exits by solid green line, while its image trajectory inside one of
the angular sector of half-aperture $\phi = \pi/3$ is shown by dashed
green line; 
\textbf{(e)}  $\mathrm{\Omega}$ is an annulus of radii $R$ and $R_c$ with an exit of half-width $\epsilon$ located on the inner radius $R$; 
\textbf{(f)}   $\mathrm{\Omega}$ is a disk and the  Brownian particle is advected by a radial
flow field $\vec{v}(r) = \mu D\; \vec{r}/r^2$, with $\mu >0$,
corresponding to an outward drift (blue arrows);
\textbf{(e)} $\mathrm{\Omega}$ is a rectangle of total width $R$ and height $2 \phi$. } 
\label{fig:1}
\end{figure} 

\begin{table}[t] 
\caption{
Summary of the results presented in this paper and comparison to
previous publications.  $T$ is the mean first passage time (MFPT) to
an exit of width $\epsilon$ from an arbitrary starting position
$\vec{r}_0$. The considered geometries are described in Fig. \ref{fig:1}.}
\label{tab:results}
 \begin{center}
 \begin{tabular}{| c | l  r | l  r |  }
  \hline 
& Previous results & &  This paper & \\[0.5ex]
& Moments &  FPT distribution  & Moments & FPT distribution \\[1ex]
   \hline
   & & &  & \\
Disk 	& $T$: exact explicit \cite{Singer:2006b,Caginalp2012}   &  $\epsilon \ll 1$,  $\vec{r}_0$ away   
	& Exact explicit    		& Exact (non-explicit)  \\
	& Averaged variance \cite{Caginalp2012} & 
	from boundaries \cite{Benichou:2010a}   & $T$, variance 		& Approximate (explicit)\\
& & 	& skewness, kurtosis 		&			\\[1ex]
Angular Sector 	
	& 	&  $\epsilon \ll 1$,  $\vec{r}_0$ away   
	& Exact explicit  	& Exact (non-explicit)\\
& & from boundaries \cite{Benichou:2010a}  & $T$, variance 	& Approximate (explicit) \\
& &	& skewness, kurtosis 	&  			\\[1ex]
Annulus & $T$: $(\epsilon, R/R_c) \ll 1 \,$ \cite{Singer2006}	& $\epsilon \ll 1$,  $\vec{r}_0$ away  
	& Approximate $T$   	& Exact (non-explicit) \\
	& 			& from boundaries \cite{Benichou:2010a} 
	& for all $R_c$ 	& Approximate (explicit) \\[1ex]
Rectangle 
	& $T$: $(\epsilon, \phi/R) \ll 1 \,$\cite{Singer2006}  &  $\epsilon \ll 1$,  $\vec{r}_0$ away 
	& Approximate $T$ & Exact (non-explicit)\\
	& Exact non-explicit \cite{Reingruber:2009a} & from boundaries \cite{Benichou:2010a} 
	& for all $R$    &  Approximate (explicit)\\[1ex]
Drift 	& $T$: $\epsilon \ll 1$ \cite{Benichou:2008}   & 
	& Approximate $T$    & Exact (non-explicit)\\
$\vec{v}(r) = \frac{\mu D}{r^2} \; \vec{r}$	& towards the exit & 
	&	for all $\mu$& Approximate (explicit) \\[1ex]
  \hline
\end{tabular}
 \end{center}
 \label{tab:summary}
\end{table}  

We apply this approach to the following domains: disks, angular sectors, annuli and rectangles (see Fig. \ref{fig:1}). Table \ref{tab:results} summarizes the
new results in this paper. For Brownian particles confined in an angular sector, we provide the
exact explicit expression for the MFPT and for the variance of the
exit time (Sec. \ref{sec:momentsdisk}).  In the case of a disk, we
obtain an expression for the Fourier coefficients of the MFPT which is
much simpler than the earlier expression from
Ref. \cite{Singer:2006b}.  We point out that the variance of the exit
time for an arbitrary starting point was previously known only through
its leading order term in the small exit limit \cite{Benichou:2010a}.
We also compute the exact skewness and excess kurtosis of the exit
time for a Brownian particle started from an arbitrary point within an
angular sector.  Away from a boundary layer near the exit, we show
that the ratio of the standard deviation to the MFPT is close to $1$,
the skewness to $2$, and the excess kurtosis to $6$, indicating that
the FPT distribution can be well approximated by an exponential
distribution, in contrast to the statement of Ref. \cite{Mattos2012}.

We exhibit the following non-intuitive result on the MFPT of Brownian particles confined to an annulus of radii $R$ and $R_c$ (Sec. \ref{sec:annuli}): under an analytically determined criteria, the MFPT is an optimizable function of the radius $R_c$.  This result is based on an
approximate expression for the MFPT which is in quantitative agreement
with numerical simulations even for a large exit size and for
arbitrary radius $R_c$.  In contrast to the
classical narrow-escape formulas for the MFPT in 2D domains of
Ref. \cite{Singer:2006c} which are not valid for degenerate domains
(in which one of dimensions is much smaller than the others), our
approximate expression of the MFPT is accurate even in the extreme
case $R_c = R$ which corresponds to a circle.  Our approximate
expressions are also accurate for rectangular confinements
(Sec. \ref{sec:rectangle}). 

In Sec. \ref{sec:drift}, we consider Brownian particles biased by a $1/r$ radial drift and confined in a disk. This situation is encountered in the biological modelling literature: the trajectories of marked proteins or tracers within the cytoplasm can be quantitatively described by an advection drift which models the effect of the intermittent active transport due
to molecular motors stochastically binding and unbinding to
microtubules \cite{Lagache:2008a}. 

In Sec. \ref{sec:heat} and Sec. \ref{sec:microchannelflow}, we explain why the FPT problem can be equivalently formulated in at
least two other physical contexts.  The first is a heat transfer
problem \cite{Carslaw1959,Crank1975} in which the temperature in a room with
adiabatic walls and an open window can be deduced from our solution
for the FPT distribution to the window (Sec. \ref{sec:heat}).  The
second is a hydrodynamic problem in which the flow rate in a
microchannel with ultra-hydrophobic walls
\cite{Sbragaglia2007,Joseph2006,Cottin-Bizonne2004} can be deduced
from our explicit expressions for the MFPT to the exit
(Sec. \ref{sec:microchannelflow}).

\section{General formalism and application to angular sectors} \label{sec:case1}

\subsection{Model and basic equations} \label{sec:basicequations}

We consider a Brownian particle confined in a bounded domain
$\mathrm{\Omega} \subset \mathbb{R}^2$, with an exit $\mathrm{\Gamma}
\subset \partial \mathrm{\Omega}$ located on an otherwise reflecting boundary
$\partial\mathrm{\Omega} \backslash \mathrm{\Gamma}$.  The
probabilistic description of this restricted diffusion relies on the
diffusive propagator $\tilde{G}^{(t)}(\vec{r}; \vec{r}_a)$, i.e., the
probability density for a particle to move from an initial position
$\vec{r} = (r, \theta)$ to a vicinity of the arrival position $\vec{r}_a$ in time
$t$, without exiting the domain.  The diffusive propagator
satisfies a backward diffusion equation 
\cite{Redner:2001a,Gardiner:2004}
\begin{equation}
\label{eq:propagator}
\frac{\partial \tilde{G}^{(t)}(\vec{r}; \vec{r}_a)}{\partial t} = D \mathrm{\Delta} \tilde{G}^{(t)}(\vec{r}; \vec{r}_a) ,
\end{equation}
where $D$ is the diffusion coefficient, and $\mathrm{\Delta}$ the
Laplace operator acting on the initial position $\vec{r}$.  The initial condition at $t = 0$ on $\tilde{G}^{(t)}(\vec{r}; \vec{r}_a)$ is given by a Dirac distribution
$\delta(\vec{r}-\vec{r}_a)$,
\begin{equation}
\label{eq:propagator_ini}
\tilde{G}^{(t=0)}(\vec{r}; \vec{r}_a) = \delta(\vec{r}-\vec{r}_a),
\end{equation}
fixes the arrival position at $\vec{r}_a$, while the mixed boundary
conditions incorporate the reflecting boundary with an absorbing exit:
\begin{align}
  \tilde{G}^{(t)}(\vec{r}; \vec{r}_a) &= 0, \quad  & \vec{r} \in \mathrm{\Gamma} ,  \label{eq:propagator_bc1} \\
  \pa_{n} \tilde{G}^{(t)}(\vec{r}; \vec{r}_a) &= 0, \quad  & \vec{r} \in \partial\mathrm{\Omega} \backslash \mathrm{\Gamma},   \label{eq:propagator_bc2}
\end{align}
where $\partial_n = \partial/\partial n$ is a shortcut notation for
the normal derivative.  The Dirichlet boundary condition
(\ref{eq:propagator_bc1}) mimics the absorbing character of the exit
$\mathrm{\Gamma}$ (i.e., the process is stopped once the particle hits
the exit), while the Neumann boundary condition
(\ref{eq:propagator_bc2}) means no flux across the remaining
reflecting boundary $\partial\mathrm{\Omega} \backslash
\mathrm{\Gamma}$.  The mixed character of the boundary conditions
presents the major challenge in solving this classical boundary value
problem.


In this paper, we consider planar domains $\mathrm{\Omega} = \{
(r,\theta)\in \R^2~:~ R_c < r < R,~ 0 < \theta < 2\phi \}$ which in
polar coordinates $(r,\theta)$ are $2\phi$-periodic along the angular
coordinate $\theta$ and bounded in the radial coordinate $r$ by $R_c$
and $R$ (e.g., disk ($\phi = \pi$) and angular sector shown in
Fig. \ref{fig:1} for $R_c = 0$).  The exit $\mathrm{\Gamma}$ is an arc
$\theta \in [\phi - \epsilon, \phi + \epsilon]$ within an otherwise
reflecting boundary $\partial\mathrm{\Omega} \backslash
\mathrm{\Gamma}$ at $r = R$.  The boundary condition $r = R_c$ is
reflecting.  Note that the exit can also be located on the inner
circle, in which case one writes $R < r < R_c$ instead of $R_c < r <
R$.
%
The angular sector geometry also accounts for the case of multiple
regularly spaced exits within a disk.  As illustrated on
Fig. \ref{fig:1}(c) for the case $n =3$, the exit through any of $n$
regularly spaced exits of width $2\epsilon$ within a disk can be
equivalently represented as the exit through (i) a single opening of
width $2\epsilon$ at the center of an angular sector of width
$2\pi/n$, or (ii) through a single opening of width $\epsilon$ in the
corner of an angular sector of width $\pi/n$.

The time needed for the particle started at $\vec{r}_0 = (r,\theta)$
to reach the exit is denoted as a random variable $\tau$.  The
survival probability up to time $t$, denoted by $\tilde{S}^{(t)}(r,
\theta)$, is the probability that the exit time $\tau$ is larger than
$t$: $\tilde{S}^{(t)}(r, \theta) = \mathbb{P}\left\{ \tau \geq t
\lvert \ X(0) = (r, \theta) \right\}$.  Since the arrival position
does not matter for the survival probability, $\tilde{S}^{(t)}(r,
\theta)$ is simply obtained by integrating the diffusive propagator
over the arrival positions $\vec{r}_a$:
\begin{equation}
\tilde{S}^{(t)}(r, \theta) = \int\limits_\mathrm{\Omega} \tilde{G}^{(t)}(r,\theta; \vec{r}_a) ~ \mathrm{d}\vec{r}_a.
\end{equation}
According to Eqs. (\ref{eq:propagator} -- \ref{eq:propagator_bc2}), the
survival probability satisfies the following equations
\cite{Redner:2001a,Gardiner:2004}
\begin{subequations}
\begin{align}
\label{eq:heat}
\frac{\partial \tilde{S}^{(t)}(r, \theta)}{\partial t} & = D \mathrm{\Delta} \tilde{S}^{(t)}(r, \theta) ,  \quad  &  (r, \theta) \in \mathrm{\Omega} \\
\label{eq:heat_ini}
\tilde{S}^{(t=0)}(r, \theta) &= 1,  \quad  & (r, \theta) \in \mathrm{\Omega}  \\
  \tilde{S}^{(t)}(r,\theta) &= 0, \quad  & r = R, \quad & \theta \in [\phi-\epsilon,\phi+\epsilon],  \label{eq:heat_bc1} \\
  \pa_{r} \tilde{S}^{(t)}(r,\theta) &= 0, \quad  & r = R, \quad & \theta \in [0,\phi-\epsilon)\cup (\phi+\epsilon,2\phi],  \label{eq:heat_bc2}  \\
  \pa_r S^{(t)}(r,\theta) &= 0 \quad  & r = R_c, \quad & \theta \in [0,2\phi).
  \label{eq:heat_b3}  \\
  \pa_\theta \tilde{S}^{(t)}(r,\theta)  &= 0 \quad  & r \in [R_c, 1], \quad & \theta \in \left\lbrace 0,2\phi \right\rbrace.  \label{eq:heat_bc4}
\end{align}
\end{subequations}
where $\partial_r = \partial/\partial r$ is a shortcut notation for
the radial derivative (note that here $\partial_r = \partial_n$), and
the Laplace operator $\mathrm{\Delta}$ in the polar coordinates is
\begin{align} \label{def:laplaceeq}
\mathrm{\Delta} = \pdds{}{r} + \frac{1}{r} \pd{}{r} + \frac{1}{r^2} \pdds{}{\theta}.
\end{align} 
The last Eq. (\ref{eq:heat_bc4}) incorporates the reflecting boundary condition at
the rays $\theta = 0$ and $\theta = 2\phi$ or, equivalently, the $2\phi$-periodicity
of the domain. \\

In this article, we study the exit time statistics through the Laplace transform of the
survival probability $\tilde{S}^{(t)}(r, \theta)$, defined for all $p
\geq 0$ as
\begin{align} \label{def:laplace}
S^{(p)}(r,\theta) \equiv \int_{0}^{\infty} \! \exp(-p \, t) \, \tilde{S}^{(t)}(r, \theta)\mathrm{d}t.
\end{align}
The FPT probability density is $\tilde{\rho}^{(t)}(r, \theta) = -
\frac{\partial \tilde{S}^{(t)}(r, \theta)}{\partial t}$.
Alternatively, one can compute $\tilde{\rho}^{(t)}(r, \theta)$ through
the inverse Laplace transform of
\begin{align} \label{eq:inverse}
\rho^{(p)}(r, \theta) \equiv 1 - p \ S^{(p)}(r,\theta).
\end{align}
The series expansion of $\exp(-pt)$ in Eq. (\ref{def:laplace}) yields
\begin{align} \label{eq:Sexpansion}
S^{(p)}(r,\theta) &= \sum^{\infty}_{n = 1} \frac{(-p)^{n-1}}{n!} \esp{\tau^n_{(r,\theta)}} , 
\end{align}
from which the $n$-th moment of the exit time is
\begin{align} \label{eq:link}
\esp{\tau^n_{(r,\theta)}} = (-1)^{n-1} \ \left[\frac{\partial^{n-1} S^{(p)}(r,\theta)}{\partial p^{n-1}}\right]_{p=0}, \qquad n \geq 1.
\end{align}
In particular, the mean FPT (MFPT) to reach the exit from a starting point $\vec{r} = (r, \theta)$ reads
\begin{align}
\esp{\tau_{(r,\theta)}} &\equiv \int^{\infty}_{0} \! t  \, \tilde{\rho}^{(t)}(r, \theta)  \,  
\mathrm{d}t= \int^{\infty}_{0} \! \tilde{S}^{(t)}(r, \theta)  \,  \mathrm{d}t = S^{(0)}(r,\theta) .
\label{eq:mfpttosurvival}
\end{align}
Last, it is useful to introduce  the global MFPT (GMFPT), denoted $\overline{\esp{\tau}}$,
as the MFPT averaged over all starting positions $\vec{r} = (r,\theta) \in
\mathrm{\Omega}$:
\begin{align} \label{def:general_definition_average}
\overline{\esp{\tau}} \equiv \frac{1}{\lvert \mathrm{\Omega} \lvert } \int_{\mathrm{\Omega}} \mathrm{d} \vec{r}\ \esp{\tau_{\vec{r}}},
\end{align}
where $\mathrm{d} \vec{r}$ is the uniform measure over
$\mathrm{\Omega}$. \\

We now consider the Laplace transform of Eq. (\ref{eq:heat} -- \ref{eq:heat_bc4}), which
yields
\begin{subequations}
\begin{align}
D \ \mathrm{\Delta} \ S^{(p)}(r,\theta)  &= p  \ S^{(p)}(r,\theta) - 1, \quad & (r, \theta) \in \mathrm{\Omega}, \label{eq:besselfull0} \\
  S^{(p)}(r,\theta) &= 0, \quad  & r = R, \quad & \theta \in [\phi-\epsilon,\phi+\epsilon],  \label{eq:bc1} \\
  \pa_{r} S^{(p)}(r,\theta) &= 0, \quad  & r = R, \quad & \theta \in [0,\phi-\epsilon)\cup (\phi+\epsilon,2\phi].   \label{eq:bc2} \\
   \pa_r S^{(p)}(r,\theta) &= 0 \quad  & r = R_c, \quad & \theta \in [0,2\phi), \label{eq:bc3}\\
\pa_\theta S^{(p)}(r,\theta)  &= 0 \quad  & r \in [R_c, R], \quad & \theta \in \left\lbrace 0,2\phi \right\rbrace. \label{eq:bc4}
 \end{align}
\end{subequations}
The Laplace transform simplifies the resolution of a heat equation (\ref{eq:heat}) into the resolution of an inhomogeneous Helmholtz equation (\ref{eq:besselfull0}). Solutions of the Helmholtz equation in cylindrical coordinates generally involve 
the modified Bessel functions of the first kind $I_n(r)$, defined as the solutions $y(x)$ of the differential equation
\begin{align}
r^2 \pdds{y}{r} + r \pd{y}{r} - (r^2 + n^2) y = 0,
\end{align}
which are finite at $r = 0$ for positive $n$. 

In the rest of this section we introduce dimensionless quantities $r
\leftarrow r/R$, $S^{(p)} \leftarrow D S^{(p)} /R^2$, and $p
\leftarrow R^2 p/D$, and define the following auxiliary function:
\begin{equation} \label{eq:auxillary}
u^{(p)}(r,\theta) \equiv S^{(p)}(r,\theta) - S^{(p)}_{\pi}(r),
\end{equation}
where $S^{(p)}_{\pi}(r)$ is the rotation invariant solution of
Eq. (\ref{eq:besselfull0}) satisfying $S^{(p)}_{\pi}(1) = 0$ and
$\pa_{r} S^{(p)}_{\pi}(r) = 0$ at $r = R_c$. 
 In the case of diffusion
inside an angular sector (with $R_c =0$), $S^{(p)}_{\pi}(r)$ can be
written in terms of the zeroth-order modified Bessel function $I_0(z)$
of the first kind as
\begin{equation} \label{def:Spi}
S^{(p)}_{\pi}(r) = \frac{1}{p} \left(1 - \dfrac{I_{0}(\sqrt{p} r)}{I_{0}(\sqrt{p})}\right)
\end{equation}
(expression for the case $R_c > 0$ is provided in Table
\ref{tab:ptable}).  Note that if the entire boundary at $r = 1$ is
absorbing (i.e., $\epsilon = \phi$), the solution of
Eq. (\ref{eq:besselfull0}) is $S^{(p)}(r,\theta) = S^{(p)}_{\pi}(r)$
\cite{Grebenkov2013}.

In terms of the auxiliary function $u^{(p)}(r,\theta)$,
Eqs. (\ref{eq:besselfull0} -- \ref{eq:bc4}) become
\begin{subequations}
\begin{align} 
 \mathrm{\Delta} u^{(p)}(r,\theta)  &= p  \ u^{(p)}(r,\theta), & (r, \theta) & \in \mathrm{\Omega} \label{eq:bessel} \\
  u^{(p)}(r,\theta) 		  &= 0 ,  & r &= 1, \quad  & \theta \in [\phi-\epsilon,\phi+\epsilon], \label{eq:bc1u}\\
 \pa_r u^{(p)}(r,\theta)  &= - \pa_r S^{(p)}_{\pi}(r), \quad  & r &= 1, \quad  & \theta \in [0,\phi-\epsilon)\cup(\phi+\epsilon,2\phi], \label{eq:bc2u} 
\\
  \pa_r u^{(p)}(r,\theta) &= 0 \quad  & r &= R_c, \quad & \theta \in [0,2\phi).
\label{eq:bc3u}
\\
\pa_\theta u^{(p)}(r,\theta)  		  &= 0 \quad  & r & \in [R_c, 1], \quad & \theta \in \left\lbrace 0,2\phi \right\rbrace, \label{eq:bc4u} 
 \end{align}
\end{subequations}
Using the separation of variables method, we express a general
solution of Eq. (\ref{eq:bessel}) which satisfies the periodicity
$\theta \rightarrow \theta + 2 \phi$ as
\begin{align}
u^{(p)}(r,\theta) &= \frac{a^{(p)} _0}{2} f^{(p)}_0(r)  + \sum^{\infty}_{n =1} a_{n}^{(p)} f^{(p)}_n(r) 
\cos \left( \frac{ n \pi \theta}{\phi}  \right), \qquad (r, \theta) \in [0, \ 1)\times[0,\phi].  \label{eq:FourierBessel}
\end{align}
where the functions $f^{(p)}_n$ depend on the considered geometry.
Since the unknown Fourier coefficients $a_n^{(p)}$ stand in front of
$f^{(p)}_n$, one can choose an appropriate normalization of the
functions $f^{(p)}_n$.  We choose the normalization condition
$f^{(p)}_n(1) = \phi/\pi$ $(n \geq 0)$.  In the case of Brownian
particles inside an angular sector ($R_c = 0$), $f^{(p)}_n$ are
expressed in terms of modified Bessel functions $I_n(z)$ of the first
kind:
\begin{align} \label{def:fn_wedge}
f^{(p)}_n(r) &= \frac{\phi}{\pi} \frac{I_{n \pi/ \phi}(\sqrt{p} r)}{I_{n  \pi/ \phi}(\sqrt{p})}, \qquad n \geq 0.
\end{align}
The function $S^{(p)}_{\pi}(r)$ defined in Eq. (\ref{eq:auxillary}) is
\begin{align}
\label{eq:spifunctionoff0}
S^{(p)}_{\pi}(r) = \frac{1 - \frac{\pi}{\phi} f^{(p)}_0(r)}{p}.
\end{align}
The Fourier coefficients $a_{n}^{(p)}$ will be uniquely determined
through the boundary conditions (\ref{eq:bc1u}) and (\ref{eq:bc2u}).
Substituting Eq. (\ref{eq:FourierBessel}) into Eqs. (\ref{eq:bc1u})
and (\ref{eq:bc2u}) leads to the system of equations
\begin{subequations}
\begin{align}
\frac{a^{(p)} _0}{2} + \sum^{\infty}_{n =1} a_{n}^{(p)} \cos \left(\frac{n\pi\theta}{\phi}\right)   
&= 0, \quad  & \theta \in [\phi-\epsilon,\phi], \label{eq:bc1complete}\\
\left[ \pa_r f^{(p)}_0 \right]_{\lvert r=1} \frac{a^{(p)} _0}{2} + \sum^{\infty}_{n =1} a_{n}^{(p)} \left[ \pa_r f^{(p)}_n \right]_{\lvert  r=1} 
\cos \left(\frac{n\pi\theta}{\phi}\right)   &= - \left[\pa_r S^{(p)}_{\pi}\right]_{\lvert  r=1}, \quad  & \theta \in [0,\phi-\epsilon), \label{eq:bc2complete}
\end{align}
\end{subequations}
where the angular coordinate $\theta$ was limited to the half-range
$[0,\phi]$ (instead of $[0,2\phi]$) due to the symmetry of these
equations with respect to the change $\theta \to 2\phi-\theta$ (this
symmetry is also related to the reflection symmetry of the domain with
respect to the ray $\theta = \phi$).

In the next section, we propose two schemes (exact and approximate) to
solve Eqs. (\ref{eq:bc1complete}) and (\ref{eq:bc2complete}).

\subsection{Resolution schemes}
\label{sec:resolution}

\subsubsection{Exact explicit expression for the MFPT in angular sector}

Let us first simplify previously known results in the case of the MFPT in a disk
($\phi= \pi$) and extend these results to angular sectors.  Using the $p\ll 1$ asymptotic expansion,
\begin{align}
\frac{I_{n}(\sqrt{p} r)}{I_{n}(\sqrt{p})} = r^n \left\lbrace 1 +\frac{\left(r^2-1\right) p}{4 (1+n)} \right\rbrace +  \mathcal{O}(p^2), \qquad n \geq 0,
\end{align}
we show that in the particular case of the MFPT ($p=0$),
Eqs. (\ref{eq:bc1complete}) and (\ref{eq:bc2complete}) read
\begin{subequations}
\begin{align}
\frac{a^{(0)}_{0}}{2} + \sum^{\infty}_{n =1} a^{(0)}_{n} \cos(n \theta)   &= 0, \quad  & \theta \in [\pi-\epsilon,\pi], \label{eq:bc1p0}\\
\sum^{\infty}_{n =1} n a^{(0)}_{n} \cos(n \theta)   &= \frac12, \quad  & \theta \in [0,\pi-\epsilon). \label{eq:bc2p0}
\end{align}
\end{subequations}
In Ref. \cite{Singer:2006b}, the solution of these equations was
provided in the form:
\begin{subequations}
\begin{align}
a^{(0)}_{0} &= \frac{\sqrt{2}}{\pi}  \int_{0}^{\pi - \epsilon} \mathrm{d}x \frac{x \sin(x/2)}{\sqrt{\cos x + \cos \epsilon }}, \label{eq:a0singer} \\
a^{(0)}_{n} &= \frac{1}{\sqrt{2} \pi} \int_{0}^{\pi - \epsilon}\mathrm{d}t \left(  \pd{}{t} \int_{0}^{t} \mathrm{d}x  \frac{x \sin(x/2)}{\sqrt{\cos x - \cos t }} 
\right) \bigl[P_n(\cos t) + P_{n-1}(\cos t) \bigr], \quad n \geq 1  ,\label{eq:ansinger}
\end{align} 
\end{subequations}
where $P_n(x)$ are Legendre polynomials.

In fact, we show in Appendix \ref{eq:a0an} that these equations can be simplified as::
\begin{subequations}
\begin{align}  \label{eq:a0careyphi_pi}
a^{(0)}_{0} &=  - 2 \ln \left[\sin \left( \frac{\epsilon}{2}\right) \right], \\
a^{(0)}_{n} &= \frac{(-1)^{n-1}}{2 n} \bigl[ P_{n}(\cos \epsilon) + P_{n-1}(\cos \epsilon) \bigr], \qquad n \geq 1. \label{eq:ancareyphi_pi}
\end{align} 
\end{subequations}

We also extend these results to an angular sector of half-aperture
$\phi$ and radius $r = 1$: $\mathrm{\Omega} = \{ (r,\theta)\in \R^2~:~
0\leq r < 1,~ 0 < \theta < 2\phi\}$.  Under the change of variables
$\hat{\theta} = \theta \pi/\phi$ and $\hat{\epsilon} = \epsilon
\pi/\phi$, Eqs. (\ref{eq:bc1complete}) and (\ref{eq:bc2complete}) are
reduced to Eqs. (\ref{eq:bc1p0}), (\ref{eq:bc2p0}) in the limit $p=0$,
from which
\begin{subequations}
\begin{align}  \label{eq:a0carey}
a^{(0)}_0 &=  \alpha_0 \equiv  - 2 \ln \left[\sin \left( \frac{\epsilon \pi}{2 \phi}\right) \right], \\
a^{(0)}_n &= \alpha_n \equiv  \frac{(-1)^{n-1}}{2 n}  \left[ P_{n}\left(\cos \frac{\epsilon \pi}{\phi}\right) + P_{n-1}\left(\cos \frac{\epsilon \pi}{\phi}\right) \right], \qquad n \geq 1.\label{eq:ancarey}
\end{align} 
\end{subequations}

We conclude that the MFPT from the angular sector for a
particle started at position $(r,\theta)$ is
\begin{align} \label{eq:1stmoment}
\esp{\tau_{(r,\theta)}}
= & \frac{1- r^2}{4} + \frac{\alpha_0}{2} \frac{\phi}{\pi}  + \frac{\phi}{\pi}
\sum^{\infty}_{n = 1}  \alpha_n ~ r^{n \pi/\phi} ~ \cos \left( \frac{n\pi \theta}{\phi} \right) ,
\end{align}
where $\phi$ is the half-aperture of the angular sector, and
$\epsilon$ is the half-width of the centered exit (see Fig. \ref{fig:1}(b)).  The GMFPT defined
in Eq. (\ref{def:general_definition_average}) reads
\begin{align} \label{def:averaged1stmoment}
\overline{\esp{\tau}} \equiv \dfrac{2}{\phi} \int^{\phi}_{0} \! \! \int^{1}_{0} r \, \mathrm{d}r \, \mathrm{d\theta} \ 
\esp{\tau^n_{(r,\theta)}} = \frac{1}{8} + \frac{\alpha_0}{2} \frac{\phi}{\pi}.
\end{align}
To our knowledge, the results in Eqs. (\ref{eq:a0carey}), (\ref{eq:ancarey}), (\ref{eq:1stmoment}) and
(\ref{def:averaged1stmoment}) are new.

Last, as described in Fig. \ref{fig:1}(d), we recall that the exit through a window of width $\epsilon$ at the corner of the sector of angle $\pi/m$ can be equivalently represented
as the exit through any of $m$ regularly spaced openings of width
$2\epsilon$ within a disk. In the limit of an
infinite number of exits $m \rightarrow\infty$, ($\phi = \pi/m
\rightarrow 0$) at a fixed ratio $\epsilon/\phi$, the MFPT of
Eq. (\ref{eq:1stmoment}) tends to the MFPT to the fully absorbing
boundary at $r =1$, as expected \cite{Nguyen2010,Berezhkovskii:2010}.

\subsubsection{Exact resolution scheme for the survival probability} \label{sec:exactresolution}

Now we solve the system of equations on the Fourier coefficients
$a^{(p)}_n$ for an arbitrary value of $p$.  We first introduce
\begin{align} \label{eq:gammataugeneral}
\gamma^{(p)}_{n} &\equiv 1-\frac{\left[ \pa_r f^{(p)}_n \right]_{r=1}}{n},  \qquad  n \geq 1, 
 \end{align}
which we use to define the following function
\begin{align}
F^{(p)}(\hat{\theta}) &\equiv - \left[\pa_r S^{(p)}_{\pi}\right]_{\lvert  r=1}  
 - \left[ \pa_r f^{(p)}_0 \right]_{\lvert r=1}  \frac{a^{(p)} _0}{2}  
+ \sum^{\infty}_{n =1}  a_{n}^{(p)} \gamma^{(p)}_{n} n  \cos(n \hat{\theta}), \qquad \theta \in [0,\pi-\hat{\epsilon}) .   \label{eq:Ftheta} 
\end{align}
For diffusion inside an angular sector, the explicit expression for
$\gamma^{(p)}_{n}$ is
\begin{align} \label{eq:gammatau}
\gamma^{(p)}_{n} &\equiv 1- \frac{\sqrt{p}}{2n} \frac{\phi}{\pi} \frac{I_{n \pi/ \phi-1}(\sqrt{p}) + I_{n \pi/ \phi+1}(\sqrt{p})}{I_{n \pi/ \phi}(\sqrt{p})},  \qquad  n \geq 1,
\end{align}
where we have used the definition (\ref{def:fn_wedge}).

Under the change of variables $\hat{\theta}= \theta \pi/\phi$ and
$\hat{\epsilon} \equiv \epsilon \pi/\phi$, Eqs. (\ref{eq:bc1complete})
and (\ref{eq:bc2complete}) read
\begin{subequations}
\begin{align}
\frac{a^{(p)}_0}{2} + \sum^{\infty}_{n =1} a_{n}^{(p)} \cos(n \hat{\theta} )  
&= 0, & \hat{\theta}  \in [\pi-\hat{\epsilon} ,\pi], \label{eq:a1Sneddon}\\
 \sum^{\infty}_{n =1} n a_{n}^{(p)} \cos(n \hat{\theta} )
 &= F^{(p)}(\hat{\theta}), & \hat{\theta} \in [0,\pi-\hat{\epsilon}),   \label{eq:a2Sneddon}
\end{align}
\end{subequations}
The problem of determining the Fourier coefficients $a_{n}^{(p)}$ from
Eqs. (\ref{eq:a1Sneddon}) and (\ref{eq:a2Sneddon}) is closely related
to the problem considered in Ref. \cite{Sneddon1966} for a given
function $F^{(p)}(\hat{\theta})$ which was independent of
$a_{n}^{(p)}$.  The crucial difference between the present case and
the case considered in Ref. \cite{Sneddon1966} is that the function
$F^{(p)}(\hat{\theta})$ defined in Eq. (\ref{eq:Ftheta}) depends on
the unknown Fourier coefficients $a_{n}^{(p)}$.  In the rest of this
section, we adapt the method of Ref. \cite{Sneddon1966} to reduce
Eqs. (\ref{eq:a1Sneddon}) and (\ref{eq:a2Sneddon}) to a linear system
of equations for the Fourier coefficients.

We first assume that for $\hat{\theta} \in [0,\pi-\hat{\epsilon})$ we
can define a function $h^{(p)}_1(t)$ such that
\begin{align} \label{eq:h1}
\frac{a^{(p)} _0}{2} + \sum^{\infty}_{n =1} a_{n}^{(p)} \cos(n \hat{\theta})   = \cos(\hat{\theta}/2) 
\int^{\pi-\hat{\epsilon}}_{\hat{\theta}} \frac{h^{(p)}_1(t)\mathrm{d}t}{\sqrt{\cos \hat{\theta} - \cos t}}.
\end{align}
Due to the invertibility of Abel's integral operator,
Eq. (\ref{eq:h1}) determines $h^{(p)}_1(t)$ uniquely for all $t \in
[0,\pi-\hat{\epsilon})$.  Using Mehler's integral representation of
Legendre polynomials,
\begin{align}
P_n(\cos t)
&= \frac{\sqrt{2}}{\pi} \int_{0}^{t} \frac{\cos\left[(n+\frac12)x\right]}{\sqrt{\cos x - \cos t }} \mathrm{d}x , \label{eq:mehler}
\end{align}
and using the absorbing condition (\ref{eq:bc1u}), we show that the
Fourier coefficients can be expressed in terms of $h^{(p)}_1(t)$:
\begin{subequations}
\begin{align} \label{eq:a0withh1}
a^{(p)} _0 &= \sqrt{2} \int^{\pi - \hat{\epsilon}}_{0} h^{(p)}_1(t)\mathrm{d}t, \\
a^{(p)} _n &= \frac{1}{\sqrt{2}} \int^{\pi - \hat{\epsilon}}_{0} h^{(p)}_1(t) \bigl[P_n(\cos t) + P_{n-1}(\cos t) \bigr] 
\mathrm{d}t, \quad n\geq 1. \label{eq:anwithh1}
\end{align}
\end{subequations}
After integration of Eq. (\ref{eq:a2Sneddon}) from $0$ to $x$,
\begin{align}
\sum^{\infty}_{n =1} a_{n}^{(p)} \sin(n x)  = \int^{x}_{0} F^{(p)}(u)\mathrm{d}u, \qquad x \in [0,\pi-\hat{\epsilon}),
\end{align}
we find that $h^{(p)}_1(t)$ satisfies the relation
\begin{align}
\int^{\pi - \hat{\epsilon}}_{0}\mathrm{d}t \ h^{(p)}_1(t) \frac{1}{\sqrt{2}} \sum^{\infty}_{n = 1} \left[ P_n(\cos t) + P_{n-1}(\cos t) \right] \sin( n x) 
 = \int^{x}_{0} F^{(p)}(u)\mathrm{d}u. \label{eq:h1sum}
\end{align}
Using the identity [see Eq. (2. 6. 31) from Ref. \cite{Sneddon1966}]
\begin{align} \label{eq:Duffy}
\frac{1}{\sqrt{2}} \sum^{\infty}_{n = 1} \bigl[ P_n(\cos t) + P_{n-1}(\cos t) \bigr] \sin( n x) = 
\frac{\cos\left(\frac{x}{2}\right) H(x-t)}{\sqrt{\cos t - \cos x}},
\end{align}
where $H(t)$ is the Heaviside distribution, we sum the series in the
left-hand side of Eq. (\ref{eq:h1sum}) to get
\begin{align}
\int^{x}_{0} \frac{h^{(p)}_1(t)\mathrm{d}t }{\sqrt{\cos t - \cos x}} = 
\frac{1}{\cos\left(\frac{x}{2}\right)} \int^{x}_{0} F^{(p)}(u)\mathrm{d}u . \label{eq:abeltype}
\end{align}
The function $h^{(p)}_1(t)$ is determined as the solution of the
Abel-type integral equation (\ref{eq:abeltype}) and reads
\begin{align}
h^{(p)}_1(t) = \frac{2}{\pi} \frac{d}{dt} \int^{t}_{0} \frac{\sin \left(\frac{x}{2} \right)\mathrm{d}x}{\sqrt{\cos x - \cos t}}
\left[ \int^{x}_{0} F^{(p)}(u)\mathrm{d}u \right],  \quad t \in [0,\pi-\hat{\epsilon}). \label{eq:h_1}
\end{align}
Substitution of Eq. (\ref{eq:h_1}) into Eqs. (\ref{eq:a0withh1}) and
(\ref{eq:anwithh1}) leads to the set of equations
\begin{subequations}
\begin{align}
a^{(p)}_0 &= \frac{2 \sqrt{2}}{\pi} \int_{0}^{\pi - \hat{\epsilon}} \mathrm{d}x \frac{\sin(x/2)}{\sqrt{\cos x + \cos \epsilon }} 
\left[\int_{0}^{x} F^{(p)}(u)\mathrm{d}u \right], \label{eq:a0pc} \\
a^{(p)}_n &= \frac{\sqrt{2}}{\pi} \int_{0}^{\pi - \hat{\epsilon}}\mathrm{d}t \left\lbrace  \pd{}{t} \int_{0}^{t} \mathrm{d}x 
\frac{\sin(x/2)}{\sqrt{\cos x - \cos t }} \left[ \int_{0}^{x} F^{(p)}(u)\mathrm{d}u \right] \right\rbrace 
\bigl[ P_n(\cos t) + P_{n-1}(\cos t) \bigr],  \quad n\geq 1. \label{eq:anpc}
\end{align}
\end{subequations}
From Eq. (\ref{eq:Ftheta}), we see that $F^{(p)}(u)$ is a linear
combination of the unknown Fourier coefficients $a^{(p)}_m$, thus
Eqs. (\ref{eq:a0pc}) and (\ref{eq:anpc}) define a linear system of
equations.  We proceed by simplifying Eqs. (\ref{eq:a0pc}) and
(\ref{eq:anpc}) in order to provide explicit relations between the
Fourier coefficients.

(i) We first simplify the identity (\ref{eq:a0pc}) using the relation
\begin{align}
2 m \alpha_m &= \frac{2 \sqrt{2}}{\pi} \int_{0}^{\pi - \hat{\epsilon}} \mathrm{d}x \frac{\sin\left( \frac{x}{2} \right) 
\sin\left(m x\right) }{\sqrt{\cos x + \cos \hat{\epsilon} }} , \qquad m \geq 1. \label{eq:Nmerased}
\end{align}
To prove Eq. (\ref{eq:Nmerased}), we express the terms $P_m(\cos
\hat{\epsilon})$ and $P_{m-1}(\cos \hat{\epsilon})$ in the
definition (\ref{eq:ancarey}) of $\alpha_m$ through the Mehler's
identity (\ref{eq:mehler}).

We substitute the explicit expression for $F^{(p)}(u)$ from
Eq. (\ref{eq:Ftheta}) into Eq. (\ref{eq:a0pc}).  Using the integral
representation of $\alpha_{m}$ from Eqs. (\ref{eq:a0singer}) and
(\ref{eq:Nmerased}), we obtain
\begin{align} \label{eq:an0tosolve}
a_{0}^{(p)} &=  - 2 \alpha_{0} \left( \left[\pa_r S^{(p)}_{\pi}\right]_{\lvert r=1} + \frac12 \left[ \pa_r f^{(p)}_0 \right]_{\lvert r=1} a^{(p)}_0\right) 
+ \sum_{m = 1}^{\infty} 2  m \gamma^{(p)}_{m} \alpha_m \ a_{m}^{(p)}.
\end{align}

(ii) We now simplify the relation (\ref{eq:anpc}) for $a^{(p)}_n$.
Substituting the explicit expression (\ref{eq:Ftheta}) for
$F^{(p)}(u)$ into Eq. (\ref{eq:anpc}) leads to the following system of
equations
\begin{align} \label{eq:matrixn}
a_{n}^{(p)} &= -2 \alpha_{n} \left(\left[\pa_r S^{(p)}_{\pi}\right]_{\lvert r=1} + \frac12 \left[ \pa_r f^{(p)}_0 \right]_{\lvert r=1} a^{(p)}_0\right)  
+ \sum_{m = 1}^{\infty} M_{nm} \gamma^{(p)}_{m}  a_{m}^{(p)} , \quad n \geq 1,
\end{align}
where the matrix $M_{nm}$ is
\begin{align} \label{eq:munsimplified}
M_{nm} &= \frac{\sqrt{2}}{\pi}  \int_{0}^{\pi  - \hat{\epsilon}} \left[  \pd{}{t} \int_{0}^{t} \mathrm{d}x  
\frac{\sin(x/2) \sin\left(m x\right)}{\sqrt{\cos x - \cos t }}  \right] \bigl[ P_n(\cos t) + P_{n-1}(\cos t) \bigr] \qquad n \geq 1, \quad  m \geq 1.
\end{align}
We show in Appendix \ref{sec:Mmatrix} that the expression for $M_{nm}$
can be simplified into
\begin{align} \label{eq:Mepsilon}
M_{nm} = \frac{m}{2} \int^{1}_{-\cos\left(\hat{\epsilon}\right)} \frac{1}{1+x} \bigl[ P_m(x)+P_{m-1}(x) \bigr] \bigl[ P_n(x)+P_{n-1}(x) \bigr] \mathrm{d}x,  
\qquad n \geq 1, \quad  m \geq 1.
\end{align}
Interestingly, the set of coefficients $\alpha_n$ is invariant under
the action of $M$: $M \cdot \alpha = \alpha $ (see Appendix
\ref{sec:Mmatrix}). 

(iii) We now write explicitly the system of equations on $a^{(p)}_n$.
We first define the set of coefficients $(\widetilde{a}^{(p)}_{n})$
defined through the following matrix inversion:
\begin{align} \label{eq:anmatrix1}
\widetilde{a}^{(p)}_{n} &\equiv \left[ \left(I - M \gamma^{(p)} \right)^{-1} \alpha\right]_n ,\qquad n \geq 1 ,
\end{align}
where $I$ stands for the identity matrix, and $\gamma^{(p)}$ is a
diagonal matrix formed by $\gamma^{(p)}_n$.  For an angular sector and
$p=0$, one has $\gamma^{(0)} = 0$ and retrieves the expected identity
$\widetilde{a}^{(0)}_{n} = \alpha_{n}$.  From
Eq. ({\ref{eq:matrixn}}), we have
\begin{align} \label{eq:anintermsofAn}
a_{n}^{(p)} &= - 2 \widetilde{a}^{(p)}_{n} \left(\left[\pa_r S^{(p)}_{\pi}\right]_{\lvert r=1} + 
\frac12 \left[ \pa_r f^{(p)}_0 \right]_{\lvert r=1} a^{(p)}_0\right), \qquad n \geq 1.
\end{align}
Substituting Eq. (\ref{eq:anintermsofAn}) into Eq. (\ref{eq:an0tosolve})
we obtain a closed system of linear equations for $a_{0}^{(p)}$:
\begin{align}  \label{eq:a0withoutC}
\nonumber
a_{0}^{(p)}\left( 
1 + \left[ \pa_r f^{(p)}_0 \right]_{\lvert r=1} \alpha_{0} \right)  &=  -2 \left[\pa_r S^{(p)}_{\pi}\right]_{\lvert  r=1} \alpha_{0}    \\
& - \left(\left[\pa_r S^{(p)}_{\pi}\right]_{\lvert  r=1} + \frac12 \left[ \pa_r f^{(p)}_0 \right]_{\lvert r=1} a^{(p)}_0\right) 
\left(  \sum_{m = 1}^{\infty}   4  m \alpha_{m} \ \widetilde{a}^{(p)}_{m} \ \gamma^{(p)}_{m} \right).
\end{align}
Introducing
\begin{align} \label{eq:Cexact}
\mathcal{C}^{(p)} \equiv \alpha_{0} + \sum_{m = 1}^{\infty}   2 m \alpha_{m} \ \widetilde{a}^{(p)}_{m} \ \gamma^{(p)}_{m} ,
\end{align}
the Fourier coefficients of the Laplace transform of the survival
probability take the compact exact form:
\begin{subequations}
\begin{align} \label{eq:a0exact}
a_{0}^{(p)} &= \mathcal{C}^{(p)} \left\lbrace \frac{- 2 \left[\pa_r S^{(p)}_{\pi}\right]_{\lvert  r=1} }{1 +
\mathcal{C}^{(p)} \left[ \pa_r f^{(p)}_0 \right]_{\lvert r=1}  } 
 \right\rbrace, \\
a_{n}^{(p)} &=  \widetilde{a}^{(p)}_{n} \left\lbrace \frac{- 2 \left[\pa_r S^{(p)}_{\pi}\right]_{\lvert  r=1} }
{1 + \mathcal{C}^{(p)} \left[ \pa_r f^{(p)}_0 \right]_{\lvert r=1}  } \right\rbrace, \qquad n \geq 1. \label{eq:anexact}
\end{align}
\end{subequations}

This solution depends on the coefficients $\widetilde{a}^{(p)}_{n}$
given by Eq. (\ref{eq:anmatrix1}). The numerical implementation of the
solution from Eqs. (\ref{eq:a0exact}) and (\ref{eq:anexact}) requires
the truncation of the matrix $M$ involved in Eq. (\ref{eq:anmatrix1})
to a finite size $N\times N$.  In spite of the truncation, we will
refer to the results obtained by this numerical procedure as exact
solutions, as their accuracy can be arbitrarily improved by increasing
the truncation size $N$ (we checked numerically that the truncation
errors decay very rapidly with $N$).  In practice, we set $N = 100$.

In the next section, we propose an approximate expression for the
Fourier coefficients $a^{(p)}_n$ which does not rely on a matrix
inversion.

\subsubsection{Approximate resolution scheme} \label{sec:approximate}

 The obtention of an approximate solution, which provides a
concise and explicit expression for the FPT, is one of the main result of this paper.
The idea of the approximate solution is to substitute the matrix $M$
by the identity matrix in Eq. (\ref{eq:anmatrix1}).  In
Refs. \cite{Benichou:2010,Benichou:2011a} and \cite{Rupprecht:2012a}, such a substitution was shown to be
efficient to compute the MFPT of a particle alternating phases of
surface and bulk diffusions in a spherically symmetric domain.

The substitution of the matrix $M$ by the identity matrix is exact for
$\epsilon = 0$ as the asymptotic expansion of $M_{nm}$ in the
limit $\epsilon \ll 1$ reads (see Appendix
\ref{sec:app_perturbatif})
\begin{align} \label{eq:Misdiagonalatfirstoder}
M_{nm} &=  \delta_{nm} + \frac{nm^2 (-1)^{n+m}}{8} \hat{\epsilon}^4 +  \mathcal{O}(\hat{\epsilon}^5), \qquad n \geq 1, \quad m \geq 1,
\end{align}
where $\delta_{nm}$ is the Kronecker symbol.  The approximation
$M_{nm} = \delta_{nm}$ allows one to invert the matrix in
Eq. (\ref{eq:anmatrix1}), yielding the following approximate solution:
\begin{align} \label{eq:Anexplicit}
\widetilde{a}^{(p)}_{n} &\approx \frac{\alpha_n}{1 - \gamma^{(p)}_n}.
\end{align}
Within the approximate scheme, we define
\begin{align} \label{eq:substitution}
\mathcal{C}_a^{(p)} &\equiv \alpha_{0} + \sum_{m = 1}^{\infty} \frac{ 2 m \alpha^2_{m}  \gamma^{(p)}_{m}}{1 -  \gamma^{(p)}_{m}}, 
\end{align}
and then substitute $\mathcal{C}^{(p)}$ by $\mathcal{C}_a^{(p)}$ in
Eqs. (\ref{eq:a0exact}) and (\ref{eq:anexact}).  Numerical simulations
indicate a $\mathcal{O}(\hat{\epsilon})$ discrepancy between the
approximate and the exact solutions.  Note that the approximate
solution is also exact in the limit $\hat{\epsilon} = \pi$, as it
predicts $\alpha_n = 0$ for all $n \geq 0$.

In the next section, we test the accuracy of the approximate
expression for the FPT distribution in the disk.  We show that the
approximate expression describes accurately the exact FPT distribution
for any value of $\hat{\epsilon}$ between $0$ and $\pi$.

\subsection{Results for the disk}

In this section, we focus on the FPT distribution for a Brownian
particle confined in the disk (Fig. \ref{fig:1}(a)).

\subsubsection{Short--time and long--time behaviour of the distribution of exit time} \label{sec:distribution}

\begin{figure}[t]
\centering
\includegraphics[width=15cm]{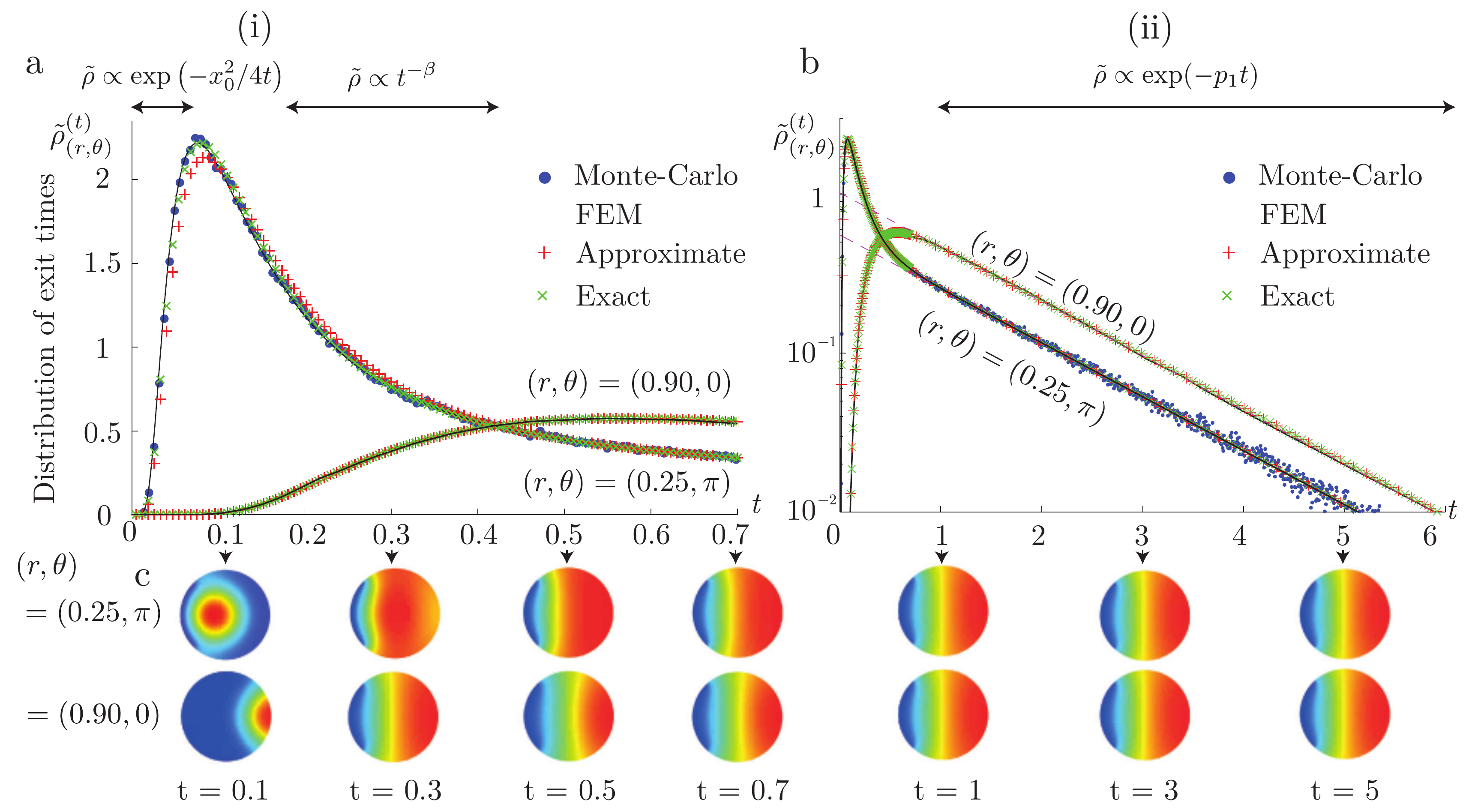} 
\caption{
(Color Online) \textbf{Upper panel}: The probability density of the
FPT to exit the unit disk through an aperture of half-width $\epsilon
= \pi/4$ (Fig. \ref{fig:1}) for Brownian particles started at $(r,
\theta) = (0.25, \pi)$ and $(r, \theta) = (0.90, 0)$.  The probability
density is plotted on (\textbf{a}) linear scale for $t \in [0 , 0.7]$
and (\textbf{b}) log-linear scale for $t \in [0 , 6]$.  The exact
solution from Eqs. (\ref{eq:a0exact}) and (\ref{eq:anexact}) (green
crosses) is compared to the approximate solution from
Sec. \ref{sec:approximate} (red pluses), finite element method (solid
black line), and Monte Carlo simulations (blue circles) (see Appendix
\ref{sec:app_montecarlo}).
\textbf{Lower panel}: (\textbf{c}) Diffusive propagator
$\tilde{G}^{(t)}(\vec{r}_0,\vec{r})$ computed by a FEM at times $t =
0.1, 0.3, 0.5, 0.7, 1, 3, 5$ for the initial positions $(r,
\theta) = (0.25, \pi)$ (top row) and $(r, \theta) = (0.90, 0)$ (bottom
row).  Color changes from dark red to dark blue correspond to changes
of the diffusive propagator from large to small values.  The FPT
probability density remains close to zero during the time needed for
the diffusive propagator to spread to the exit.  After a time $t >
R^2/D = 1$, (i) the diffusive propagator reaches a steady state
profile, and (ii) the FPT is close to an exponential distribution with
the decay rate constant $p_1$ predicted by Eq. (\ref{eq:longtime}) and
shown by magenta dashed lines in \textbf{(b)}. }
\label{fig:2}
\end{figure}

\begin{figure}[t]
\centering
\includegraphics[width=15cm]{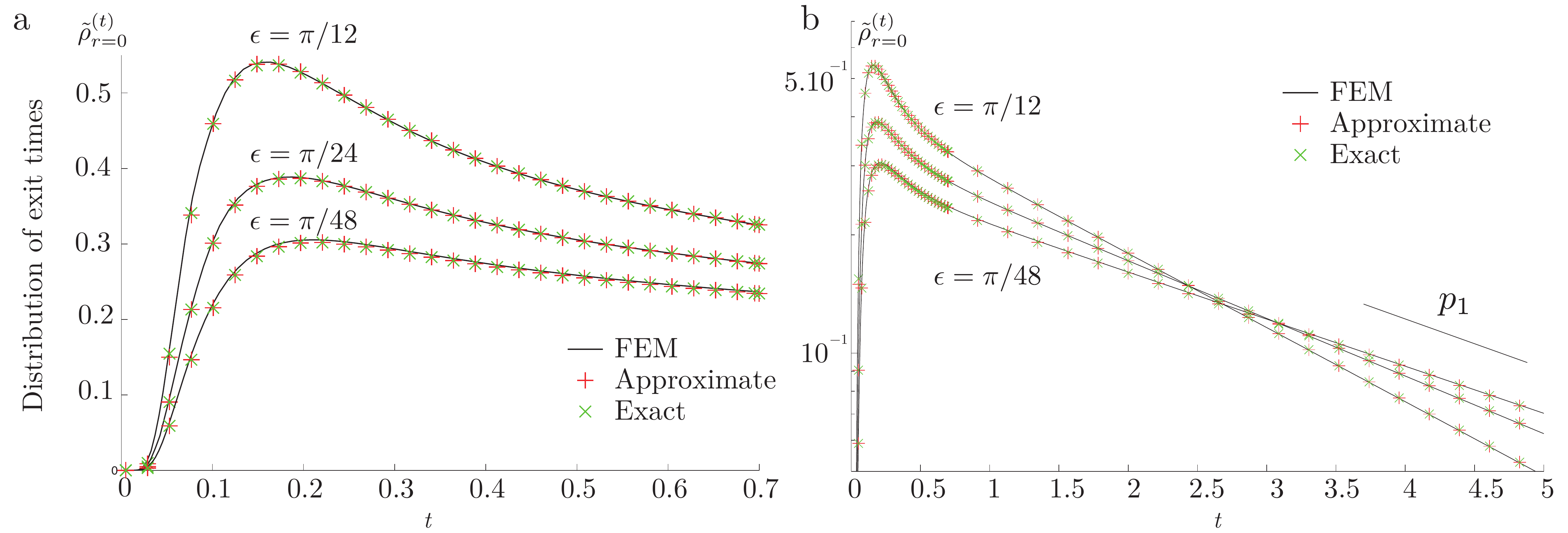} 
\caption{
(Color Online) The FPT probability density for a disk with an exit of
half-width $\epsilon = \pi/12, \pi/24, \pi/48$ for Brownian particles
started at $r =0$, in (\textbf{a}) linear scale for $t \in [0 , 0.7]$,
and (\textbf{b}) log-linear scale for $t \in [0 , 5]$.  The exact
solution from Eqs. (\ref{eq:a0exact} -- \ref{eq:anexact}) (green
crosses) is compared to its analytical approximation from
Sec. \ref{sec:approximate} (red pluses), and a finite element method
resolution (black solid line), showing an excellent agreement (units:
$R^2/D = 1$). }
\label{fig:2b} 
\end{figure}

Figures \ref{fig:2} and \ref{fig:2b} show the exact and approximate
FPT probability densities that are computed through the inverse
Laplace transform of $\rho^{(p)}(r, \theta)$ from  Eqs. (\ref{eq:inverse}),  (\ref{eq:a0exact}) and (\ref{eq:anexact}).  The exact and approximate
solutions agree well with the numerical results which are obtained by
two independent techniques: (i) a finite element method (FEM)
resolution of Eqs. (\ref{eq:heat} -- \ref{eq:heat_bc2}) in the time domain
by COMSOL \cite{RogerW.Pryor2009}, and (ii) Monte Carlo simulations of a large
sample of random walks (see Appendix \ref{sec:app_montecarlo} for
further information on these computational techniques).  In both
numerical solutions, the problem is solved in time domain, i.e.,
without Laplace transform inversion.  Even for a large exit size
$\epsilon = \pi/4$, the approximate solution from
Sec. \ref{sec:approximate} agrees well with both the exact solution and 
the numerical results (see Fig. \ref{fig:2}).  The agreement is improved uniformly in time for
smaller values of $\epsilon$ (Fig. \ref{fig:2b}).

The time scale $R^2/D$ separates the short-time and the long-time
behaviors of the FPT distribution, as illustrated on Figs. \ref{fig:2}(a--b).
In addition, Fig. \ref{fig:2}(c) shows the diffusive propagator
$\tilde{G}^{(t)}(\vec{r}_0,\vec{r})$ as a function of the arrival
position $\vec{r}$ (computed by a finite element method in COMSOL).
The spatial distribution of the diffusive propagator provides the following
physical insight on the evolution of the FPT in the short-time (i) and
long-time regimes (ii): \\

(i) For $t \ll R^2/D$, most particles do not have enough time to reach
the confining boundary (\textit{a fortiori} the exit) and the
confining domain appears to be almost infinite.  This can be seen on
the profile of the diffusive propagator for $t < 1$ in
Fig. \ref{fig:2}(c).  If the distance $x_0$ from the initial
position $(r, \theta)$ to the center of the exit $(1, \pi)$, $x_0(r,
\theta) = \sqrt{1+r^2-2r \cos(\theta)}$, is sufficiently
small (e.g., for $(r, \theta) = (0.25, \pi)$ in Fig. \ref{fig:2}), the
short-time behavior of the FPT probability density is approximatively
\begin{align} \label{sec:onedFPT}
\tilde{\rho}^{(t)}(r, \theta) \approx \frac{x_0(r,\theta)}{\sqrt{4\pi D t^3}} \exp\left(-\frac{x^2_0(r, \theta)}{4 D t}\right),
\end{align}
which describes the FPT for a particle started at distance $x_0$ from
the absorbing endpoint of a semi-infinite segment \cite{Redner:2001a}.
The exponential factor strongly dominates at short times when
$\sqrt{4Dt} \ll x_0(r,\theta)$.  At the intermediate times, when
$x_0(r,\theta) \ll \sqrt{4Dt} \ll R$, the FPT probability density
exhibits a power law decay, $\tilde{\rho}\propto t^{-\beta}$, with
$\beta = 3/2$.  For a starting position $(r, \theta) = (0.90, \pi)$, a fit
of the FPT probability density at the intermediate times $t \in [0.1 ,
0.7]$ by a power law distribution $t^{-\beta}$ yields $\beta = 1.49 \pm 0.02$, a
value which is close to the corresponding value of $\beta$ in a
semi-infinite system ($\beta = 1.50$, see Eq. (\ref{sec:onedFPT})).  \\

(ii) For $t \gg R^2/D$, the FPT probability density exhibits an
exponential tail.  The terms in braces in Eqs. (\ref{eq:a0exact}) and
(\ref{eq:anexact}) determine the decay rate of the exponential tail.
Note that in Fig. \ref{fig:2} the rescaled profile of the diffusive
propagator appears stationary for all $t > R^2/D = 1$, as expected.  \\

The survival probability $\tilde{S}^{(t)}$ (resp. the FPT probability
density $\tilde{\rho}^{(t)}$) can be expressed as the sum over the
residues of the Laplace transform $S^{(p)}$ (resp. $\rho^{(p)}$).  For
example, if the boundary is fully absorbing (i.e., $\epsilon = \pi$),
the survival probability of a particle started at $(r, \theta)$ can be written from Eq. (\ref{def:Spi}).
Indeed, after a spectral decomposition of the Laplace operator with Dirichlet boundary condition on the disk \cite{Grebenkov2013},
the coefficients can be computed from the residue theorem applied to Eq. (\ref{def:Spi}), leading to:
\begin{align} \label{eq:wedgepi}
\tilde{S}^{(t)}(r, \theta)= \sum^{\infty}_{k = 1} \frac{2}{\xi_{0k}} 
\frac{J_0\left( \xi_{0k} r \right)}{ J_1\left( \xi_{0k} \right)} \exp\left(- \xi^2_{0k} t\right),
\end{align}
where the coefficients $-\xi_{0k}$ are the poles of $S^{(p)}_{\pi}(r)$
(as a function of $p$), as the coefficients  $\xi_{0k}$ are
the zeros of the zeroth order Bessel function: $J_0(\xi_{0k}) = 0$ for
all $k \geq 1$.  Note that the functions $S^{(p)}$ and $\rho^{(p)}$
are related through Eq. (\ref{eq:inverse}) and therefore have the same
poles.  In the general case $\epsilon < \pi$, the long-time behavior
of the FPT probability density is governed by the smallest decay rate
$p_1$: $\tilde{\rho}^{(t)}_{\pi}(r)$ asymptotically decays as
$\exp(-p_1 t)$ for $t \gg R^2/D$.  The quantity $-p_1$ is the largest negative root of the equation
\begin{align}
1 + \mathcal{C}^{(p_1)} \left[ \pa_r f^{(p_1)}_0 \right]_{\lvert r=1} = 0.
\label{eq:longtime}
\end{align}
The latter equation (\ref{eq:longtime}) uniquely determines $p_1$ and can be solved
numerically (see Fig. \ref{fig:3a}).  Note that
Eq. (\ref{eq:longtime}) is independent of the starting position of the
particle: in the long-time limit, particles have lost memory of their
starting positions.  
In the next section, we provide explicit estimates of $p_1$, which yield the long time asymptotics of the FPT distribution in the narrow-escape limit $\epsilon \ll 1$ and beyond.


\subsubsection{Beyond the narrow--escape limit: a simplified expression for the long-time decay rate} \label{sec:p1approximate}

\begin{figure}[t] 
\centering
\includegraphics[width=14cm]{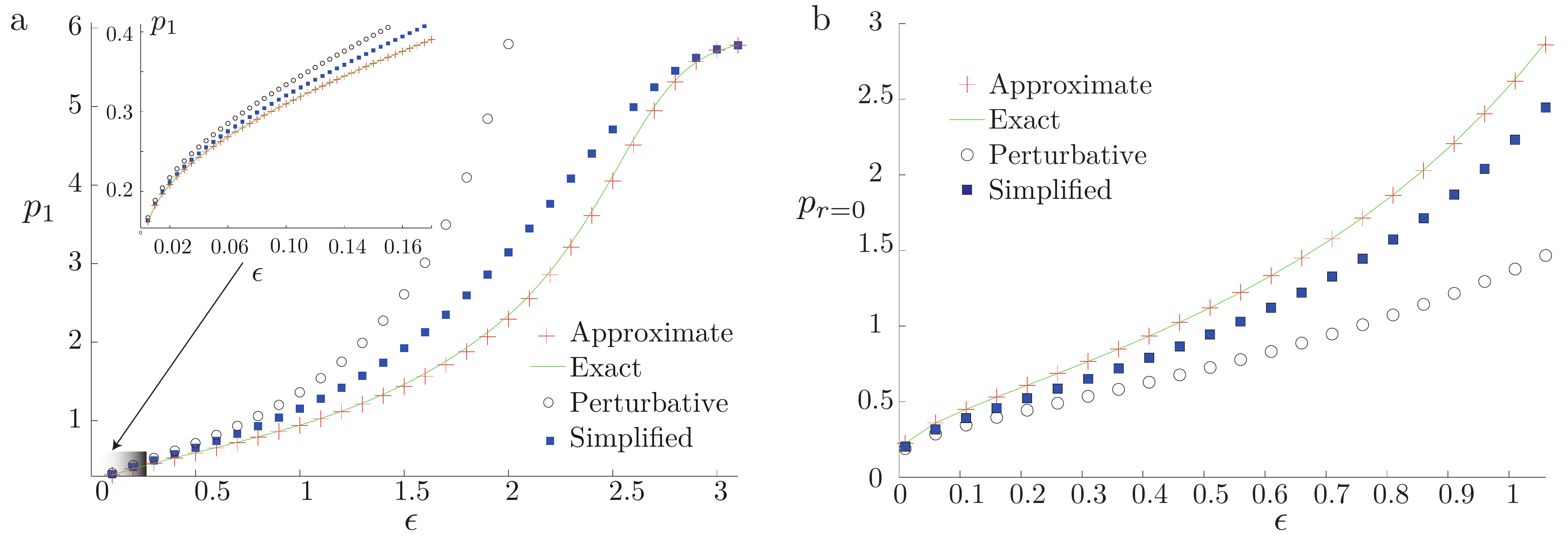} 
\caption{
(Color Online) Comparison of three approximate schemes describing the
long-time behavior of the survival probability for a Brownian particle
started at $r = 0$ which exits a confining disk through an exit of
half-width $\epsilon$.  The quantities $p_{r=0}$ and $p_1$ are defined
from the asymptotic expansion: $\log(\tilde{\rho}^{(t)}(r = 0)) \simeq
\log(p_{r=0}) - p_1 t$ according to Eq. (\ref{eq:firstorderterm}).
\textbf{(a)} Decay rate $p_1$ of the long-time limit of the survival
probability as a function of the exit half-width $\epsilon$.  The rate
$p_1$ is obtained through: an exponential interpolation of the exact
distribution from Eqs. (\ref{eq:a0exact}) and (\ref{eq:anexact})
(solid green line), an exponential interpolation of the approximate
distribution from Sec. \ref{sec:approximate} (red pluses), the
asymptotic expression $p_1 = 2/\alpha_0$ in the limit $\epsilon \ll 1$
\cite{Benichou:2010a} (black circles), and simplified
Eq. (\ref{eq:simplified}) (blue squares).  Note that
Eq. (\ref{eq:simplified}) provides accurate results in both limits
$\epsilon = 0$ and $\epsilon = \pi$, and is more accurate than the
perturbative expansion.  \textbf{(b)} The prefactor $p_{r=0}$ to the
exponential distribution defined in Eq. (\ref{eq:firstorderterm}) as a
function of the exit half-width $\epsilon$.  }
\label{fig:3a}
\end{figure}

The determination of the FPT distribution for arbitrary $\epsilon$ presented above  is the main result of the present paper. In this paragraph we first compare our result to the previously known results on the FPT distribution in the narrow--escape limit from Ref. \cite{Benichou:2010a}.  We then propose a simplified expression for $p_1$ which does not depend on the specific shape of the domain $\mathrm{\Omega}$. This simplified expression is asymptotically exact in the limit $\epsilon \ll 1$ and is in fact in good agreement with the exact expression for $p_1$ (computed through Eq. (\ref{eq:longtime}))  over the whole range of value of $\epsilon$  (see Fig. \ref{fig:3a}).

We first point out that at the first order in $\epsilon
\ll 1$, Eqs. (\ref{eq:a0carey}) and (\ref{eq:ancarey}) read
\begin{subequations}
\begin{align} 
\alpha_{0} &= 2  \ln\left( \frac{2 \phi}{\pi \epsilon}\right) + \mathcal{O}\left( \left(\frac{\epsilon\pi}{\phi}\right) ^2\right),  \label{eq:a0firstorder}\\
\alpha_{n} &= \frac{(-1)^{n-1}}{n}
 + \mathcal{O}\left( \left(\frac{\epsilon\pi}{\phi}\right) ^2\right),
 \qquad n \geq 1 .\label{eq:anfirstorder} \end{align}
\end{subequations}
The logarithmic singularity of Eq.  (\ref{eq:a0firstorder}) is a
well-known result discussed in Ref. \cite{Singer:2006b} for $\phi =
\pi$ and in Ref. \cite{ward} for $\phi < \pi$.

In this limit $\epsilon \ll 1$ and if the starting position $\vec{r}$ is located away from the frontier of the confining domain $\mathrm{\Omega}$, it has been shown in Ref. \cite{Benichou:2010a}  that the FPT converges to an exponential distribution with mean the GMFPT $\overline{\esp{\tau}}$, defined in Eq. (\ref{def:general_definition_average}). Hence Ref. \cite{Benichou:2010a}  implies the asymptotic identity: $p_1 = 1/\overline{\esp{\tau}}$, for $\epsilon \ll 1$. Due to the divergence of $\alpha_0$ from Eq. (\ref{eq:a0firstorder}), the latter identity is equivalent to:
 \begin{align} \label{eq:perturbativep1}
p_1 = \frac{\pi}{\lvert \mathrm{\Omega} \lvert} \frac{2}{\alpha_0}, \quad  \forall \epsilon \ll 1.
\end{align}
where $\lvert \mathrm{\Omega} \lvert$ stands for the volume of $\mathrm{\Omega}$. We stress that the latter expression in Eq. (\ref{eq:perturbativep1})  depends  on $\lvert \mathrm{\Omega} \lvert$, but not on the precise shape of the domain $\mathrm{\Omega}$. This statement holds however only in the limit $\epsilon \ll 1$, since the exact  result of Eq. (\ref{eq:longtime}), which is valid for any $\epsilon$,   depends a priori on the specific geometry of the domain through the set $(\gamma^{(p)}_m)$ in the expression of $C^{(p)}$ (see Eq. (\ref{eq:Cexact})).

In fact one can propose a simple approximate expression for $p_1$ with larger range of validity in $\epsilon$ than the asymptotic relation from Eq. (\ref{eq:perturbativep1}). Let us first notice that at the leading order in $\epsilon \ll 1$, $\mathcal{C}^{(p)} \approx \alpha_0$. The latter identity leads us to substitution  $\mathcal{C}^{(p)}$ for $\alpha_0$ in Eq. (\ref{eq:longtime}).  Setting $q_1 = i \sqrt{p_1}$,
the simplified expression of Eq. (\ref{eq:longtime}) is reduced to:
\begin{align} \label{eq:simplified}
\left[ \pa_r f^{(q_1)}_0 \right]_{\lvert r=1}  \approx \frac{\pi}{\lvert \mathrm{\Omega} \lvert}  \frac{1}{q_1 \alpha_0}.
\end{align}
In the limit $\epsilon \ll 1$, the simplified expression of Eq. (\ref{eq:simplified}) leads to the perturbative result of Eq. (\ref{eq:perturbativep1}). Note that the simplified expression of
Eq. (\ref{eq:simplified}) is also exact for $\epsilon = \pi$, in which case
$\alpha_0$ tends to zero and $p_1$ tends to $\xi^2_{01}$, in agreement
with Eq. (\ref{eq:wedgepi}). 

Finally, for a given value of the decay rate $p_1$, the residue theorem leads
to the following long-time exponential decay of the survival
probability:
\begin{align}
\tilde{S}^{(t)}(r, \theta) &\approx \left(-2\left[\pa_r S^{(p_1)}_{\pi}\right]_{\lvert  r=1} \right) 
\left[ \frac{J_0\left(\sqrt{p_1} r \right)}{ J_0\left(\sqrt{p_1}\right)} 
\frac{\alpha_0}{2} + \sum^{\infty}_{m = 1} \frac{J_m\left( \sqrt{p_1} r \right)}{ J_m\left( \sqrt{p_1} \right)} 
\alpha_m \cos\left(\frac{m\theta \phi}{\pi}\right)   \right] 
\exp\left(- p_1 t\right) \\
& \approx p_{(r,\theta)} \exp\left(- p_1 t\right), \label{eq:firstorderterm}
\end{align}
where $p_{(r,\theta)}$ is the prefactor of the exponential
distribution, which depends on the   starting position  $X(0) = (r,\theta)$.  In
Fig. \ref{fig:3a}, we show that the simplified solution of
Eq. (\ref{eq:simplified}) provides a good approximation of $p_1$ over the whole range of values for $\epsilon$.

\subsection{Moments and cumulants} \label{sec:moments} 

In this section we derive exact expressions for the moments of the
exit times for a general set of functions $f^{(p)}_n, \ n \geq 0$.  We
emphasize that these expressions are fully explicit in the case of
Brownian particles confined in an angular sector.


We use the notations $a_{n}^{[k]}$ for the $k$-th coefficient in the
small $p \ll 1$ expansion of $a_{n}^{(p)}$:
\begin{align}
a_{n}^{(p)} &\equiv \sum^{\infty}_{k = 0}  p^{k} a_{n}^{[k]} , \qquad n \geq 0.
\end{align}
By definition $a^{(0)}_{n} = a_{n}^{[0]}$. Similarly we define for all
$k \geq 0$ the set of coefficients $S^{[k]}(r,\theta)$, $\left[\pa_r
S^{[k]}_{\pi}\right]_{\lvert r=1}$, $\left[\pa_r
f^{[k]}_{0}\right]_{\lvert r=1}$, $f^{[k]}_{n}$, and $\gamma^{[k]}_{n}$
for all $n \geq 0$.  From Eqs. (\ref{eq:auxillary}) and
(\ref{eq:FourierBessel}), the coefficient $S^{[j]}(r,\theta)$ is given
in terms of $a_{0}^{[j]}, \ j \geq 0$:
\begin{align} 
S^{[j]}(r,\theta) &= S^{[j]}_{\pi}(r) + \sum^{j}_{k = 0} \frac{a^{[k]} _0}{2} f^{[j -k]}_0(r)   + 
\sum^{\infty}_{n =1} \left( \sum^{j}_{k = 0} a^{[k]}_n f^{[j -k]}_n(r)  \right) \cos  \left( \frac{ n \pi \theta}{\phi}  \right).  \label{eq:FourierBessel_ordre1}
\end{align}
In the next section, we explain how the coefficients $a_{n}^{[j]}$ can
be expressed through the lower-order terms $a^{[k]}_n, 0 \leq k
\leq j - 1$.

\subsubsection{Recurrence relation on the Fourier coefficients} \label{sec:recurrence}

We show that the Fourier coefficients satisfy a hierarchical set of
equations, i.e., it is possible to express $a^{[j]}_n$ in terms of the
lower-order coefficients $a^{[k]}_n, n \geq 0$, with $k = 0,1,\ldots,
j - 1$.  The Fourier coefficients of the MFPT are obtained by setting
$p = 0$ in Eqs. (\ref{eq:a0exact}) and (\ref{eq:anexact}):
\begin{align}
\esp{\tau_{(r,\theta)}} = S^{(0)}_{\pi}(r) + \frac{a^{(0)} _0}{2} f^{(0)}_0(r)  + \sum^{\infty}_{n =1} a_{n}^{(0)} f^{(0)}_n(r) 
\cos \left( \frac{ n \pi \theta}{\phi}  \right), \qquad (r, \theta) \in \mathrm{\Omega}.  \label{eq:mfpt_general}
\end{align}
For instance, one retrieves the exact explicit expression
(\ref{eq:1stmoment}) for the MFPT of Brownian particles confined in an
angular sector.  In other geometries considered in
Sec. \ref{sec:2ddomains}, the exact resolution scheme requires a
numerical solution of linear Eqs. (\ref{eq:anmatrix1}) at $p = 0$.

According to Eq. (\ref{eq:matrixn}), the unknown coefficients
$a^{[j]}_n$ are related to the unknown coefficients $a^{[j]}_0$ and to
the known lower-order coefficients $a^{[k]}_n$, $k = 1,2,\ldots, j-1$,
\begin{align} \label{eq:recurrencean}
\sum^{\infty}_{m = 1} (\delta_{nm} - M_{nm} \gamma^{[0]}_{m})a^{[j]}_m = - 2 \alpha_n \left( \left[\pa_r S^{[j]}_{\pi}\right]_{\lvert  r=1}  
+ \frac12 \sum^{j}_{k = 0} a^{[j-k]}_0 \left[\pa_r f^{[k]}_{0}\right]_{\lvert  r=1}  \right)  + \sum^{\infty}_{m = 1} M_{nm} 
\left( \sum^{j}_{k = 1} \gamma^{[k]}_{m} a^{[j-k]}_m\right) .
\end{align}
In terms of the vector $\widetilde{\alpha}^{(0)}_n$ defined by
Eq. (\ref{eq:anmatrix1}) with $p=0$, the matrix $\widetilde{M}$ is
defined as
\begin{align} 
\widetilde{M} &\equiv (I - M \cdot \gamma^{(0)})^{-1} \cdot M.
\label{eq:Mtilde}
\end{align}
where $I$ stands for the identity matrix, and $\gamma^{(0)}$ is a
diagonal matrix formed by $\gamma^{(0)}_n$.  In terms of the matrix $\widetilde{M}$, Eq. (\ref{eq:recurrencean})
takes the form
\begin{align} \label{eq:recurrencean2}
a^{[j]}_n = - 2 \widetilde{\alpha}^{(0)}_n \left( \left[\pa_r S^{[j]}_{\pi}\right]_{\lvert  r=1}  
+ \frac12 \sum^{j}_{k = 0} a^{[j-k]}_0 \left[\pa_r f^{[k]}_{0}\right]_{\lvert  r=1}  \right)  
+ \sum^{\infty}_{m = 1} \widetilde{M}_{nm} \left( \sum^{j}_{k = 1} \gamma^{[k]}_{m} a^{[j-k]}_m\right).
\end{align}
Substituting this expression into Eq. (\ref{eq:an0tosolve}) leads to
\begin{align} \label{eq:a0orderj}
a^{[j]}_0 = \frac{- 2 \alpha_0 \left( \left[\pa_r S^{[j]}_{\pi}\right]_{\lvert  r=1}  + 
\frac12 \sum^{j}_{k = 1} a^{[j-k]}_0 \left[\pa_r f^{[k]}_{0}\right]_{\lvert  r=1} \right)  +  
\sum^{\infty}_{m = 1} 2  m \alpha_m T^{[j]}_m}{1 + \left[\pa_r f^{[0]}_{0}\right]_{\lvert  r=1} 
\left( \alpha_0 + \sum^{\infty}_{m =1} 2  m \alpha_m \widetilde{\alpha}^{(0)}_m \gamma^{[0]}_m \right) } , 
\end{align}
where
\begin{align} \label{eq:tm}
T^{[j]}_m  =  \sum^{j}_{k = 1} \gamma^{[k]}_{m} a^{[j-k]}_m +
\gamma^{(0)}_m \left[ - 2 \widetilde{\alpha}^{(0)}_m \left( \left[\pa_r S^{[j]}_{\pi}\right]_{\lvert  r=1}  + 
\frac12 \sum^{j}_{k = 0} a^{[j-k]}_0 \left[\pa_r f^{[k]}_{0}\right]_{\lvert  r=1}  \right) +
 \sum^{\infty}_{l = 1}  \widetilde{M}_{ml}  \left( \sum^{j}_{k = 1}  \gamma^{[k]}_l a^{[j-k]}_l \right) \right] .
\end{align}
Equation (\ref{eq:a0orderj}) expresses $a^{[j]}_0$ in terms of the
known coefficients $a^{[k]}_n$, $k = 1,2,\ldots, j - 1$.  The
coefficients $a^{[j]}_n~ (n \geq 1)$ are then determined through
Eq. (\ref{eq:recurrencean2}).

Following the idea of Sec. \ref{sec:resolution}, we define an
approximate scheme in which the matrix $M$ is replaced by the identity
matrix in Eqs. (\ref{eq:a0exact}), (\ref{eq:anexact}) and
(\ref{eq:Mtilde}).  This approximation leads to
Eq. (\ref{eq:Anexplicit}) and solves Eq. (\ref{eq:Mtilde}) as
\begin{align}
\widetilde{M}_{nm} &\approx \frac{M_{nm}}{1 - \gamma_n^{(0)}} .
\end{align}
We recall that the approximation $M_{mn} = \delta_{mn}$ is exact in
the limit $\epsilon = 0$ (see Eq. (\ref{eq:Misdiagonalatfirstoder})).

In the next section, we focus on Brownian particles confined in an
angular sector, in which case the recursive method of
Eq. (\ref{eq:a0orderj}) provides an exact explicit expression for the
variance and an exact computation scheme of the third and fourth moments.

\subsubsection{Diffusion in angular sector: explicit exact moments} \label{sec:momentsdisk}

\begin{figure}[t]
\centering
\includegraphics[height=12cm]{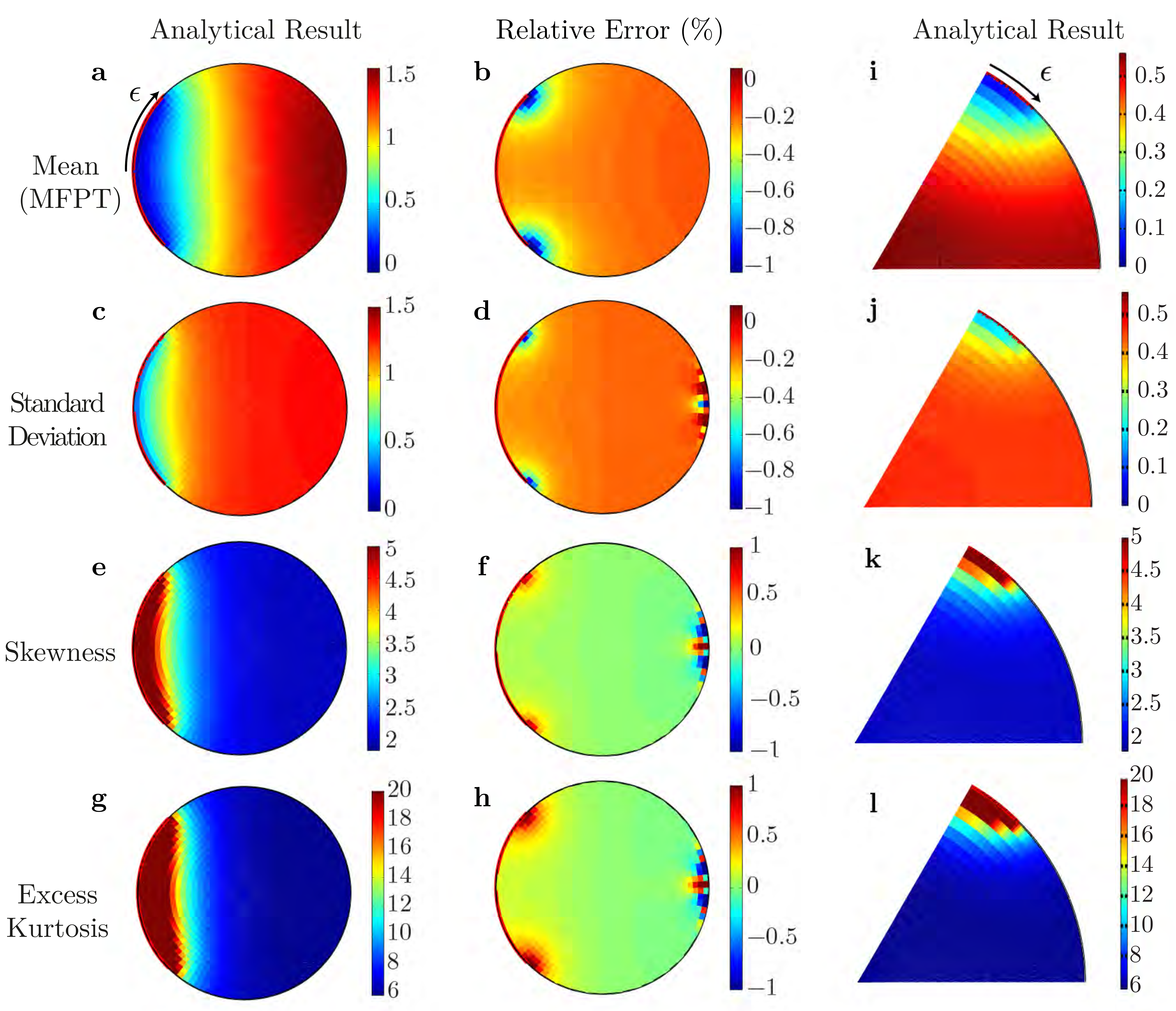}
\caption{
(Color Online) Mean, variance, skewness and excess kurtosis of the FPT
for a Brownian particle confined in: a disk ($\phi = \pi$) with the
exit width $2 \epsilon = \pi/2$ shown by red line (\textbf{left
column}), and an angular sector ($\phi = \pi/3$) with an exit at the
corner of the width $\epsilon = \phi/4$ shown by red line
(\textbf{right column}).  The middle column
(\textbf{b}-\textbf{d}-\textbf{f}-\textbf{h}) shows the relative error
$(X_n - X_a)/X_n$ between the analytical result $X_a$ and a finite
element method resolution $X_n$.  The exit of half-width $\epsilon =
\pi/4$ is shown by red line.  The relative error for the cumulants is the
largest near the edges of the exit.  Outside a boundary layer near to
the exit, the first four moments are very close to those of an
exponential distribution (for which the standard deviation is equal to
the mean, while the skewness and excess kurtosis are equal to $2$ and
$6$ respectively).  }
\label{fig:3}
\end{figure}

For Brownian particles confined in an angular sector, the coefficients
$\gamma^{(0)}_n$ are equal to zero.  The recursive scheme provides
thus an exact explicit expression for the moments of the exit time as the
resolution of Eqs. (\ref{eq:anmatrix1}) and (\ref{eq:Mtilde}) is
straightforward.  Following the method of Sec. \ref{sec:recurrence},
we obtain the second moment of the exit time by combining
Eqs. (\ref{eq:1stmoment}) and (\ref{eq:a0orderj}):
\begin{align} 
\esp{\tau^2_{(r,\theta)}}
= &\left[ \frac{1}{2} \left( \frac{\alpha_0 \pi}{\phi} \right)^2 +\frac{1}{8} \left( \frac{\alpha_0 \pi}{\phi} \right) \left(3 - 2 r^2\right) 
+ \frac{3 + r^4 - 4 \ r^2}{32}  + \sum_{m = 1}^{\infty} \frac{\phi}{m \pi +\phi} \left( \frac{\alpha_{m} \pi}{\phi} \right)^2\right] \nonumber  \\ 
&+ \sum^{\infty}_{n=1} \left[\alpha_n \left( \frac{1}{4} +  \alpha_0 \frac{\phi}{\pi} + \frac{\phi(1-r^2)}{2(n \pi+\phi)} \right) +  
\sum^{\infty}_{m = 1} \frac{M_{nm} \phi^2}{m \pi (m \pi+\phi)} \alpha_m  \right] \frac{\phi \, r^{n \pi/\phi}}{\pi}  
\cos\left(\frac{n\pi\theta}{\phi}\right). \label{eq:2ndmoment}
\end{align}
Substracting the square of the MFPT defined in
Eq. (\ref{eq:1stmoment}), we obtain the variance of the exit time for
any starting position within the angular sector.  We point out that
the variance was previously known only in the narrow-escape limit
$\epsilon \ll 1$ through its leading order term $\alpha_0^{2}/4$
\cite{Benichou:2010a}.  Figures \ref{fig:3} and \ref{fig:4} show the standard
deviation, defined as the square root of the variance, as a function
of the starting position $(r, \theta)$ within the disk ($\phi = \pi$)
and an angular sector ($\phi = \pi/3$), respectively.

The average of Eq. (\ref{eq:2ndmoment}) over all starting positions
within the angular sector (defined in Eq. (\ref{def:averaged1stmoment})) leads
to
\begin{align}  \label{eq:variance1storder}
\overline{\esp{\tau^2}} &= \frac{1}{2}\left( \frac{\alpha_0 \pi}{\phi} \right)^2  + 
\frac{1}{4} \left( \frac{\alpha_0 \pi}{\phi} \right) + \frac{1}{24} + \sum_{m = 1}^{\infty} \frac{1}{m\pi/\phi +1} \left( \frac{\alpha_{m} \pi}{\phi} \right)^2.
\end{align}
On the other hand, the spatial average of Eq. (\ref{eq:1stmoment})
turns out to be
\begin{align} \label{eq:averagesquaremfpt}
\overline{\esp{\tau}^2}  = \frac12 \overline{\esp{\tau^2}} .
\end{align}
Combining these two results, one gets the following expression of the
spatial average of the variance:
\begin{align}  \label{eq:variancefulldisk}
\overline{\Var \left[ \tau \right]} &= \overline{\esp{\tau^2}} - \overline{\esp{\tau}^2} = \overline{\esp{\tau}^2} 
= \frac{1}{4}\left( \frac{\alpha_0 \pi}{\phi} \right)^2 + \frac{1}{8} \left( \frac{\alpha_0 \pi}{\phi} \right) + \frac{1}{48} +
 \frac12 \sum_{m = 1}^{\infty} \frac{1}{m\pi/\phi +1} \left( \frac{\alpha_{m} \pi}{\phi} \right)^2.
\end{align}
The equality between the averaged variance and the averaged second
moment, $\overline{\Var \left[ \tau \right]} =
\overline{\esp{\tau}^{2}}$, was previously obtained from general
arguments \cite{Caginalp2012}.

We now consider the random variable $\tau_{\mathrm{\Omega}}$, defined
as the exit time of a particle started at a random starting position,
with uniform distribution within $\mathrm{\Omega}$.  Although the
averaged moments are identical, $\overline{\esp{\tau^n}} =
\esp{\tau_\mathrm{\Omega}^n}$ (see Appendix \ref{sec:spatialaverage}), the
variance of $\tau_{\mathrm{\Omega}}$,
\begin{align} \label{eq:varianceoverlinetau}
\Var \left[ \tau_{\mathrm{\Omega}} \right] &= \esp{\tau_{\mathrm{\Omega}}^2} - \esp{\tau_{\mathrm{\Omega}}}^2  = 
\frac{1}{4}\left( \frac{\alpha_0 \pi}{\phi} \right)^2  + \frac{1}{8} \left( \frac{\alpha_0 \pi}{\phi} \right) + 
\frac{5}{192} + \sum_{m = 1}^{\infty} \frac{1}{m\pi/\phi +1} \left( \frac{\alpha_{m} \pi}{\phi} \right)^2.
\end{align}
is different from the spatially averaged variance of
Eq. (\ref{eq:variancefulldisk}).

Following the method of Sec. \ref{sec:recurrence}, we compute the Fourier coefficient of the third moment from the Fourier coefficients of the two first moments. Similarly, we compute the fourth moment from the three first moments. We define the skewness ${\rm{Ske}}\left[\tau_{(r, \theta)}\right] $
and the excess kurtosis ${\rm{Kur}}\left[\tau_{(r, \theta)}\right]$ as
\begin{align} \label{eq:skeandkur}
{\rm{Ske}}\left[\tau_{(r, \theta)}\right] = \esp{\left(\frac{\tau - \esp{\tau} }{\sqrt{\esp{\tau^2} - \esp{\tau}^2}} \right)^3} 
\quad {\rm{and}} \quad 
{\rm{Kur}}\left[\tau_{(r, \theta)}\right] =  \esp{\left(\frac{\tau - \esp{\tau} }{\sqrt{\esp{\tau^2} - \esp{\tau}^2}} \right)^4} - 3.
\end{align}
Figures \ref{fig:3} and \ref{fig:4} show the skewness and the excess
kurtosis for a Brownian particle confined in a disk and an angular
sector, respectively.  The lower bounds for the skewness and kurtosis
are respectively $2$ and $6$, e.g. the values of the skewness and
excess kurtosis of an exponential distribution.  A positive skewness
indicates that the distribution of exit times is always skewed to the
right of the MFPT.  The distribution is also leptokurtic, meaning that
the excess kurtosis is positive: very long residence times occur more
frequently than predicted by a Gaussian distribution.

The ratio of the standard deviation to the mean, the skewness, and the
excess kurtosis diverge when the distance between the starting
position and the center of the exit tends to zero.  This is consistent
with the short-time behavior of the FPT distribution.

\subsubsection{Moments in the narrow-escape limit: the case of the disk}  \label{sec:narrowescape}

The argument of Ref. \cite{Benichou:2010a} holds in the narrow-escape limit $\epsilon \ll 1$ and for a starting position $\vec{r}$ away from the frontier of the confining domain $\Omega$. Under these two assumptions, one expects the set $\left(\esp{\tau_{(r, \theta)}^n}\right), n \geq 1$ to converge to the set $\left(n! \, \esp{\tau_{(r, \theta)}}\right), n \geq 1$ which are the set of moments of an exponential distribution of mean $\esp{\tau_{(r, \theta)}}$.

\begin{figure}[h]
\centering
\includegraphics[height=10cm]{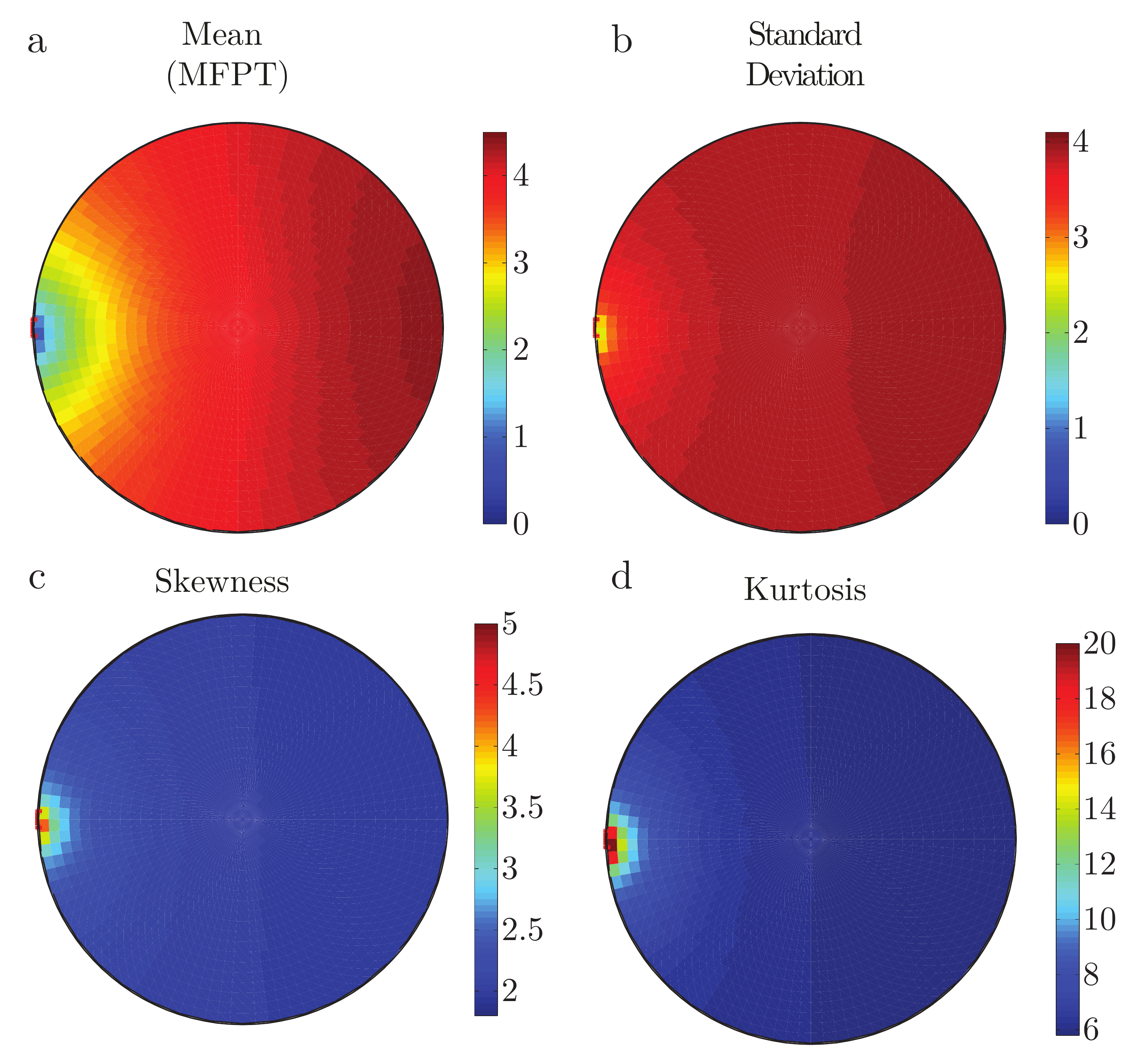}
\caption{
(Color Online) First four cumulants of the exit time from a disk for
$\epsilon = \pi/60$: (\textbf{a}) MFPT; (\textbf{b}) standard
deviation; (\textbf{c}) skewness; (\textbf{d}) excess kurtosis.  The
dashed line shows the boundary layer, i.e. the region enclosed within
the disk of radius $2
\epsilon \ln\left(\epsilon/2\right)$.  Note that outside the boundary
layer, (i) the standard deviation is approximatively equal to the
MFPT, and (ii) the skewness and excess kurtosis are approximatively
equal to $2$ and $6$ that correspond to an exponential distribution.
}
\label{fig:4}
\end{figure}

Figure \ref{fig:4} shows the first four cumulants of the exit time in
the case $\epsilon = \pi/60$ in the boundary of an unit disk.  Following the
argument of Ref. \cite{Singer:2006b}, we introduce
\begin{align}
\label{eq:boundary}
\delta = - \epsilon \ln\left(\frac{\epsilon}{2} \right),
\end{align}
and define the boundary layer $\mathcal{B}[(1, \pi), \delta]$ as the
intersection of $\mathrm{\Omega}$ with the disk of radius $2 \delta$ centered on the exit $(r, \theta) = (1,
\pi)$ (i.e. the area enclosed by the dashed line in Fig. \ref{fig:4}).
For a starting position $(r, \theta) \in \mathrm{\Omega}\backslash
\mathcal{B}[(1, \pi), \delta]$ outside the boundary layer, the ratio
of the standard deviation to the MFPT is close to $1$, while the
skewness and excess kurtosis are respectively close to $2$ and $6$, as
for an exponential distribution.  As a consequence, for a sufficiently
small exit and for a starting position $(r, \theta) \in
\mathrm{\Omega}\backslash \mathcal{B}[(1, \pi), \delta]$ outside the
boundary layer, the exit time follows approximately an exponential
distribution whose mean is the MFPT defined by
Eq. (\ref{eq:1stmoment}). This observation extends the predictions of Ref. \cite{Benichou:2010a}.


\section{Extensions and applications} \label{sec:2ddomains}

In this section, we discuss various extensions and applications of our
approach.  First, we apply the general framework of
Sec. \ref{sec:case1} to consider the exit time for a Brownian particle
from annuli (Sec. \ref{sec:annuli}).  Second, we extend the method of
Sec. \ref{sec:case1} to obtain the FPT distribution for particles
moving according to radial advection-diffusion (Sec. \ref{sec:drift}).
We also show the applicability of this method to the FPT problems in
rectangles (\ref{sec:rectangle}).  Finally, we briefly discuss the
analogies of the FPT problem to heat transfer (Sec. \ref{sec:heat})
and to microchannel flows (Sec. \ref{sec:microchannelflow}).  Table
\ref{tab:ptable} summarizes explicit expressions of the functions
$S^{(p)}_{\pi}(r)$ and $f^{(p)}_n$ for each considered domain.  Using
Table \ref{tab:ptable} to compute the Fourier coefficients in
Eqs. (\ref{eq:a0exact}) and (\ref{eq:anexact}), one gets the FPT
distribution for each considered geometry.  In addition, the recursive
scheme in Eq. (\ref{eq:a0orderj}) provides all the moments of the exit
time.

\begin{table}[h!] \label{}
\caption{
Summary of the quantities involved in the computation of the Laplace
transform of the survival probability in five studied geometries.  The
functions $f_n$ are defined for all $n \geq 0$ while the functions
$\gamma_n$ are defined for all $n \geq 1$.}
\label{tab:ptable}
 \begin{center}
 \begin{tabular}{| c | c c|}
  \hline
Case & Quantity & Series expansion in $p \ll 1$ \\
  \hline
Full disk & $S^{(p)}_{\pi}(r) = \frac{1}{p} \left(1 - \dfrac{I_{0}[\sqrt{p} r]}{I_{0}[\sqrt{p}]}\right)$ 
	& $\frac{1-r^2}{4} +\frac{\left(-3+4 r^2-r^4\right)}{64} p +\mathcal{O}(p^2)$\\
(no bias) & $f^{(p)}_n(r) = \frac{I_{n}(\sqrt{p} r)}{I_{n}(\sqrt{p})}$  
	& $r^n \left(1 +\frac{\left(r^2-1\right) p}{4 (1+n)} \right)+  \mathcal{O}(p^2)$\\
Sec. \ref{sec:case1} & $\gamma^{(p)}_n = 1 - \frac{\sqrt{p}}{2n} \frac{I_{n-1}(\sqrt{p}) + I_{n+1}(\sqrt{p})}{I_{n}(\sqrt{p})}$ 
	& $-\frac{p}{2 n (n +1)}+ \mathcal{O}(p^2)$ \\
  \hline
Angular Sector & $S^{(p)}_{\pi}(r) = \frac{1}{p} \left(1 - \dfrac{I_{0}[\sqrt{p} r]}{I_{0}[\sqrt{p}]}\right)$ 
	& $\frac{1-r^2}{4} +\frac{\left(-3+4 r^2-r^4\right)}{64} p +\mathcal{O}(p^2)$ \\
of half-width $\phi$ & $f^{(p)}_n(r) = \frac{\phi}{\pi} \frac{I_{n \pi/\phi}(\sqrt{p} r)}{I_{n \pi/\phi}(\sqrt{p})} $ 
	& $ \frac{\phi \, r^{\frac{n \pi }{\phi }}}{\pi} \left( 1 +\frac{\left(r^2-1\right) \phi  p}{4 (n \pi +\phi )} \right)+\mathcal{O}(p^2)$\\
Sec. \ref{sec:case1}  &$\gamma^{(p)}_{n} = 1-\frac{\sqrt{p}}{2n} \frac{I_{n \pi/\phi-1}(\sqrt{p}) + I_{n \pi/\phi+1}(\sqrt{p})}{I_{n \pi/\phi}(\sqrt{p})}$ 
	& $ -\frac{p \phi}{2 n \pi \left( n \frac{\pi}{\phi} +1 \right)} + \mathcal{O}(p^2)$ \\
  \hline
Full Disk & $S^{(p)}_{\pi}(r)  = \frac{1}{p} - \frac{r^{-\frac{\mu}{2}}}{p} \frac{I_{\mu/2}(\sqrt{p} r)}{I_{\mu/2}(\sqrt{p})}$ 
	& $(1-r^2) \left(\frac{1}{2(2 + \mu)}  + p \, \frac{\left(-6-\mu+\left(2+\mu\right) r^2\right)}{8 \left(2+\mu\right)^2 
	\left(4+\mu\right)} \right) + \mathcal{O}(p^2)$\\
with bias: & $f^{(p)}_n(r) = r^{-\frac{\mu}{2}} \frac{I_{\mu_n}(\sqrt{p} r)}{I_{\mu_n}(\sqrt{p})}$ 
	& $r^{(\mu_n-\mu)/2} \left( 1 + p \, \frac{r^2-1}{2 \left(2+\mu_n\right)} \right) + \mathcal{O}(p^2)$ \\
$\vec{v}(r) = \frac{\mu D}{r^2} \; \vec{r}$ & where $\mu_n = \sqrt{n^2 + \left(\frac{\mu}{2}\right)^2}$ & \\
 &  & \\
Sec. \ref{sec:drift} & $\gamma^{(p)}_n = 1-\frac{\left[ \pa_r f^{(p)}_n \right]_{r=1}}{n}$ 
	& $ 1 + \frac{\mu}{2 m} - \sqrt{1+\left(\frac{\mu}{2 m}\right)^2} -\frac{p}{2m \left( 1+\sqrt{m^2+\left(\mu/2\right) ^2} \right)}$ \\
  \hline  
  Annuli & $\rho_n(\sqrt{p} r) \equiv \frac{I_{n}[\sqrt{p} r]}{I_{n}[\sqrt{p}]}, \quad \nu_n(\sqrt{p} r) \equiv  \frac{K_{n}[\sqrt{p} r]}{K_{n}[\sqrt{p}]}$ 
& \\
 & & \\
Sec. \ref{sec:drift} &
$S^{(p)}_{\pi}(r)  = \frac{1}{p} - \frac{1}{p} \frac{\rho_0(\sqrt{p} r) 
- \frac{\rho'_0(\sqrt{p} r)}{\nu'_0(\sqrt{p} R_c)} \nu_0(\sqrt{p} r)}{1 - \frac{\rho'_0(\sqrt{p} r)}{\nu'_0(\sqrt{p} R_c)}}$ 
	&  $\frac{1}{4} \left(1-r^2+2 R_c^2 \log(r)\right)  +  \mathcal{O}(p)$\\
& $f^{(p)}_n(r) = \frac{\rho_n(\sqrt{p} r) - \frac{\rho'_n(\sqrt{p} R_c)}{\nu'_n(\sqrt{p} R_c)} \nu_n(\sqrt{p} r)}{1 - 
\frac{\rho'_n(\sqrt{p} R_c)}{\nu'_n(\sqrt{p} R_c)}}$ 
	& $\frac{r^{2 n} + Rc^{2 n}}{r^{n} (1 + Rc^{2 n})}  +  \mathcal{O}(p^2)$\\
& $\gamma^{(p)}_n = 1-\frac{\left[ \pa_r f^{(p)}_n \right]_{r=1}}{n}$ 
	& $\frac{2 Rc^{2n}}{1 + Rc^{2n}} + \mathcal{O}(p)$ \\
  \hline
Rectangle &  $S^{(p)}_{\pi}(r)  = \frac{1}{p} \left( 1- \frac{\cosh(\sqrt{p} r)}{\cosh\left(\sqrt{p} R\right)} \right)$
	& $\frac{R^2 -r^2}{2} +\frac{1}{24} \left(-r^4+6 r^2 R^2-5 R^4\right) p+  \mathcal{O}(p^2)$\\
of width $\phi = \pi$ 	& $f^{(p)}_n(r) = \frac{\cosh\left(\sqrt{p + n^2} \ r\right)}{\cosh(\sqrt{p + n^2} R)}$ 
	& $ \frac{\cosh(n r)}{\cosh(n R)}  +  \mathcal{O}(p)$ \\
Sec. \ref{sec:rectangle} & $\gamma^{(p)}_n = 1 -\frac{\sqrt{p + n^2} \tanh\left(\sqrt{p} R\right)}{n}$ & 
	$(1-\tanh(n R))-\frac{\left(\frac{n R}{\cosh(n R)^2}+\tanh(n R)\right) p}{2 n^2}+\mathcal{O}(p^2)$  \\
  \hline
\end{tabular}
 \end{center}
\end{table}

\subsection{Annuli} \label{sec:annuli}
  
We consider the confining domain $\mathrm{\Omega}$ to be an annulus
with concentric circular boundaries at $r = R = 1$ and $r = R_c$:
$\mathrm{\Omega} = \{ (r,\theta)\in\R^2~:~ R_c < r < 1,~ 0 \leq \theta
< 2\pi\}$ for $R_c < 1$ (the exit is located on the outer boundary) or
$\mathrm{\Omega} = \{ (r,\theta)\in\R^2~:~ 1 < r < R_c ,~ 0 \leq
\theta < 2\pi\}$ for $R_c > 1$ (the exit is located on the inner
boundary).  The boundary at $r = R_c$ is fully reflecting, while the
boundary at $r = 1$ is reflecting except for an absorbing arc of
length $2 \epsilon$, as illustrated in Fig. \ref{fig:annulus}(b).  The
survival probability in the Laplace space satisfies the Helmholtz
Eq. (\ref{eq:besselfull0}), the mixed Neumann-Dirichlet boundary
conditions of Eqs. (\ref{eq:bc1})-(\ref{eq:bc2}) at $r =1$ and the
Neumann boundary condition at $r = R_c$ from Eq. (\ref{eq:bc3}).

\subsubsection{Distribution of the first passage time}

\begin{figure}[h!]
\centering
\includegraphics[width=14cm]{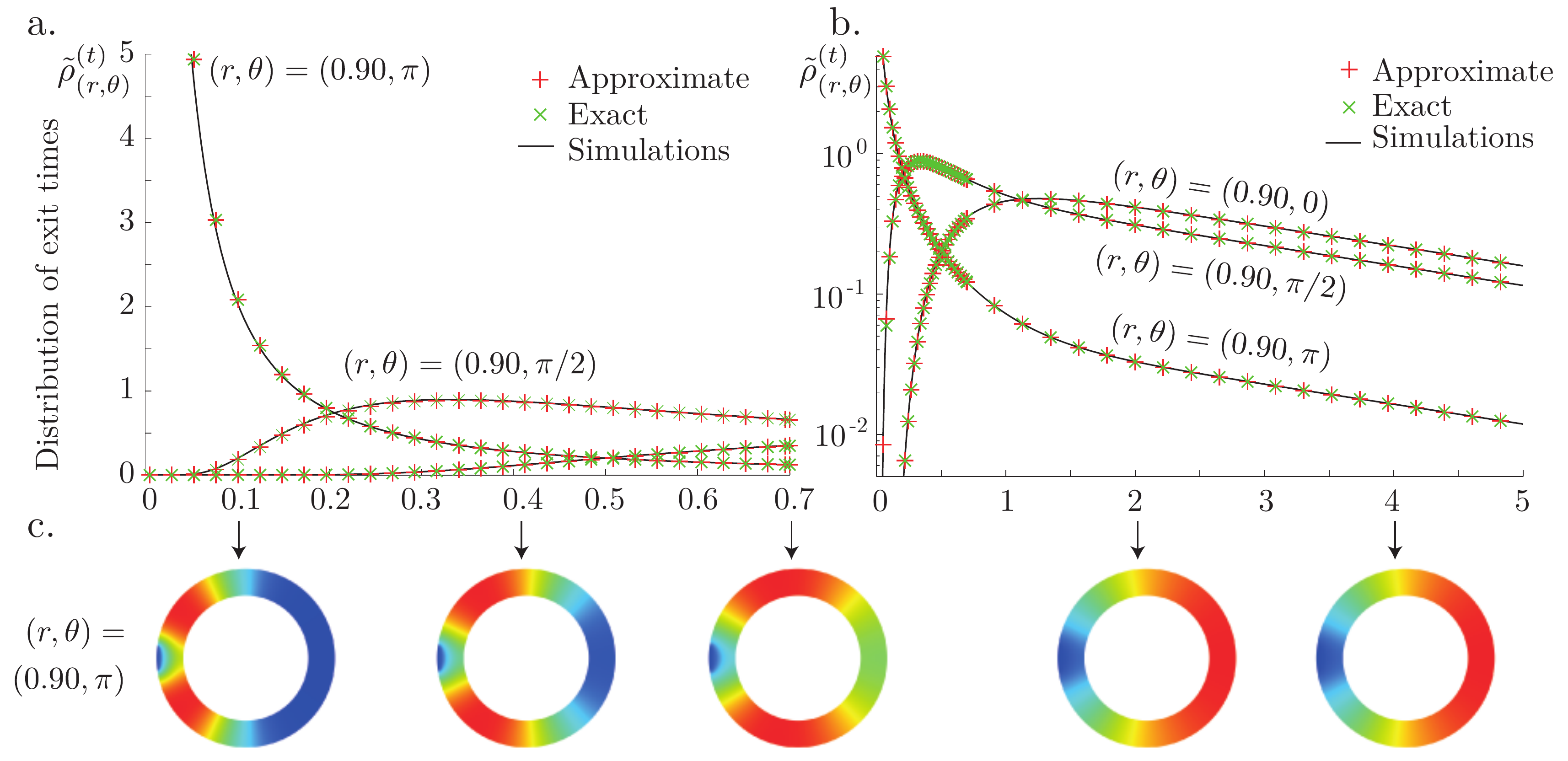} 
\caption{
(Color Online) \textbf{Upper panel}: The probability density of the
FPT to exit an annulus between two circles of radii $R_c = 0.7$ and $R
= 1$ through an exit of half-width $\epsilon = \pi/24 \approx 0.13$
within the outer radius $R$, for a Brownian particle started at
$(r,\theta) = (0.90, \pi), (0.90, \pi/2), (0.90, 0)$.  The FPT
probability density is shown in (\textbf{a}) linear scale for $t \in
[0 , 0.7]$, and (\textbf{b}) log-linear scale for $t \in [0 , 6]$.
The exact solution from Eqs. (\ref{eq:a0exact}) and (\ref{eq:anexact})
(green crosses) is compared to its analytical approximation from
Sec. \ref{sec:approximate} (red pluses) and a finite element method
numerical solution (black solid line).
\textbf{Lower panel}: Diffusive propagator
$\tilde{G}^{(t)}(\vec{r}_0,\vec{r})$ computed by a FEM at times $t =
0.1, 0.4, 0.7, 2, 4$ for the initial position $(r, \theta) = (0.90,
\pi)$.  Color changes from dark red to dark blue correspond to changes
of the diffusive propagator from large to small values.  After a time
$t > R^2/D = 1$, the diffusive propagator reaches a steady state
profile, and the FPT distribution agrees well with an exponential
distribution with the decay rate constant $p_1$.}
\label{fig:pdfann}
\end{figure}

In Fig. \ref{fig:pdfann} we represent the FPT probability density
$\tilde{\rho}^{(t)}(r, \theta)$ for an annulus with $R_c = 0.70$ and
an exit of half-size $\epsilon = \pi/24$, with three starting
positions: $(r, \theta) = (0.90, \pi)$, $(0.90, \pi/2)$, and $(0.90,
0)$.  The exact, approximate and numerical schemes agree well in the
whole range of times.

The short-time behavior of the FPT distribution strongly depends on
the initial position of the particle.  If the starting position is far
from the exit [e.g. $(r, \theta) = (0.90, 0)$], the FPT probability
density is negligible up to time $t \approx R^2/D =1$.  For a starting
position that is within the boundary layer defined in
Sec. \ref{sec:momentsdisk} [e.g. $(r, \theta) = (0.90, \pi)$], the FPT
probability density is sharply picked at $t \approx x_0^2/D =1$, where
$x_0$ is the distance to the center of the exit.  In contrast, the
long-time behavior ($t \gg R^2/D$) of the FPT probability density is
independent of the initial position.
 
In the next section, we obtain an explicit approximate expression for
the MFPT.  Using this approximate expression we find that the MFPT is
an optimizable function of $R_c$ under analytically determined
criteria.

\subsubsection{Approximate expression for the MFPT}

We substitute $\mathcal{C}^{(0)}$ by $\mathcal{C}_a^{(0)}$ from
Eq. (\ref{eq:substitution}) into Eqs. (\ref{eq:a0exact}) and
(\ref{eq:anexact}) and use the expressions of Table
\ref{tab:ptable} to get the following approximation for the
Fourier coefficients of the MFPT:
\begin{align} \label{eq:a0_annulus}
a_{0}^{(0)} &\approx  \left(1 - R_c^2\right)\left[ \alpha_0 + \sum^{\infty}_{k = 1} \left(\frac{2 R_c^{2 k}}{ R_c^{2 k}-1} \right) 2k \alpha^2_k \right]
, \\
a_{n}^{(0)} &\approx \left(1+R_c^{2 n}\right) \frac{1 - R_c^2}{1-R_c^{2 n}} \alpha_n, \quad n \geq 1, \label{eq:an_annulus}
\end{align}
where $\alpha_n$ are defined by Eqs. (\ref{eq:ancarey}).  Using
Eqs. (\ref{eq:a0_annulus}) and (\ref{eq:an_annulus}), the MFPT can be
computed from Eq. (\ref{eq:mfpt_general}).  The GMFPT defined in
Eq. (\ref{def:general_definition_average}) is
\begin{align} \label{eq:averagedmfptrc}
\overline{\esp{\tau}} \approx  \frac{1 - R_c^2}{2}\left[ \alpha_0 + \sum^{\infty}_{k = 1} \left(\frac{R_c^{2 k}}{R_c^{2 k}-1} \right) 4k \alpha^2_k \right] 
+ \frac{1}{8} \left(1-3 R_c^2\right) + \frac{1}{2}  \frac{ R_c^4 \ln(R_c)}{R_c^2 -1}.
\end{align}
Notice that for $R_c = 0$, we retrieve the exact
Eqs. (\ref{eq:1stmoment}) and (\ref{def:averaged1stmoment}).  Let us
now compare the approximate solution to previously known results in
two limits $R_c \to 1$ and $R_c \gg 1$.  This comparison provides an
error estimate of the approximate solution in the limit $\epsilon \ll
1$.

(i) In the limit $R_c \rightarrow 1$, using the identity 
\begin{align} \label{eq:sumam0}
\sum^{\infty}_{m=1} 2 m \alpha^2_m = \alpha_0,
\end{align} 
which is valid for any value of $\epsilon$ (and proved in Appendix
\ref{sec:app_sum}), we show that the coefficients $a_{n}^{(0)}$ from
Eqs. (\ref{eq:a0_annulus}) and (\ref{eq:an_annulus}) become
\begin{align} \label{eq:anRc}
a_{0}^{(0)} &\approx  2 \frac{\pi^2}{3},  \\
a_{n}^{(0)} &\approx \frac{2 (-1)^{n-1}}{n^2}, \qquad n \geq 1,
\end{align}
hence the approximate expression for the MFPT (defined in
Eq. (\ref{eq:mfpt_general})) is
\begin{align} \label{eq:mean_1D}
\esp{\tau_{(r, \theta)}}=  \frac{1}{2} (\pi - \theta) (\pi + \theta) +  \mathcal{O}(1-R_c), \qquad \theta \in \left[0 , \pi - \epsilon \right].
\end{align}
The approximate expression is equal to the MFPT of a purely
one-dimensional process up to $\mathcal{O}(\epsilon)$ terms.  The
average of Eq. (\ref{eq:mean_1D}) over all angles $\theta \in \left[0,
\pi\right]$ is equal to $\pi^3/(3 \pi)$, which is close to the exact
result $(\pi-\epsilon)^3/(3 \pi)$.  Note that asymptotic formulas of
Ref. \cite{Singer:2006b} on the MFPT in 2D domains are not valid in
the limit $R_c \rightarrow 1$.

(ii) In the large volume limit $R_c \gg 1$, we use
Eq. (\ref{eq:sumam0}) to show that the first Fourier coefficient of
the MFPT reads
\begin{align} \label{eq:a0Rc1}
a_{0}^{(0)} &\approx  \left(R_c^2 - 1\right) \left( \alpha_0 +  4 R_c^{-2} \right) + \mathcal{O}(1/R^4_c).
\end{align}
In the limit $\epsilon \ll 1$, the latter expansion can be identified
with the result of Ref. \cite{Singer:2006c} (p. 503) which is shown to
be exact up to a $\mathcal{O}(\epsilon)$ term.

\begin{figure}[t]
\centering
\includegraphics[height=7.5cm]{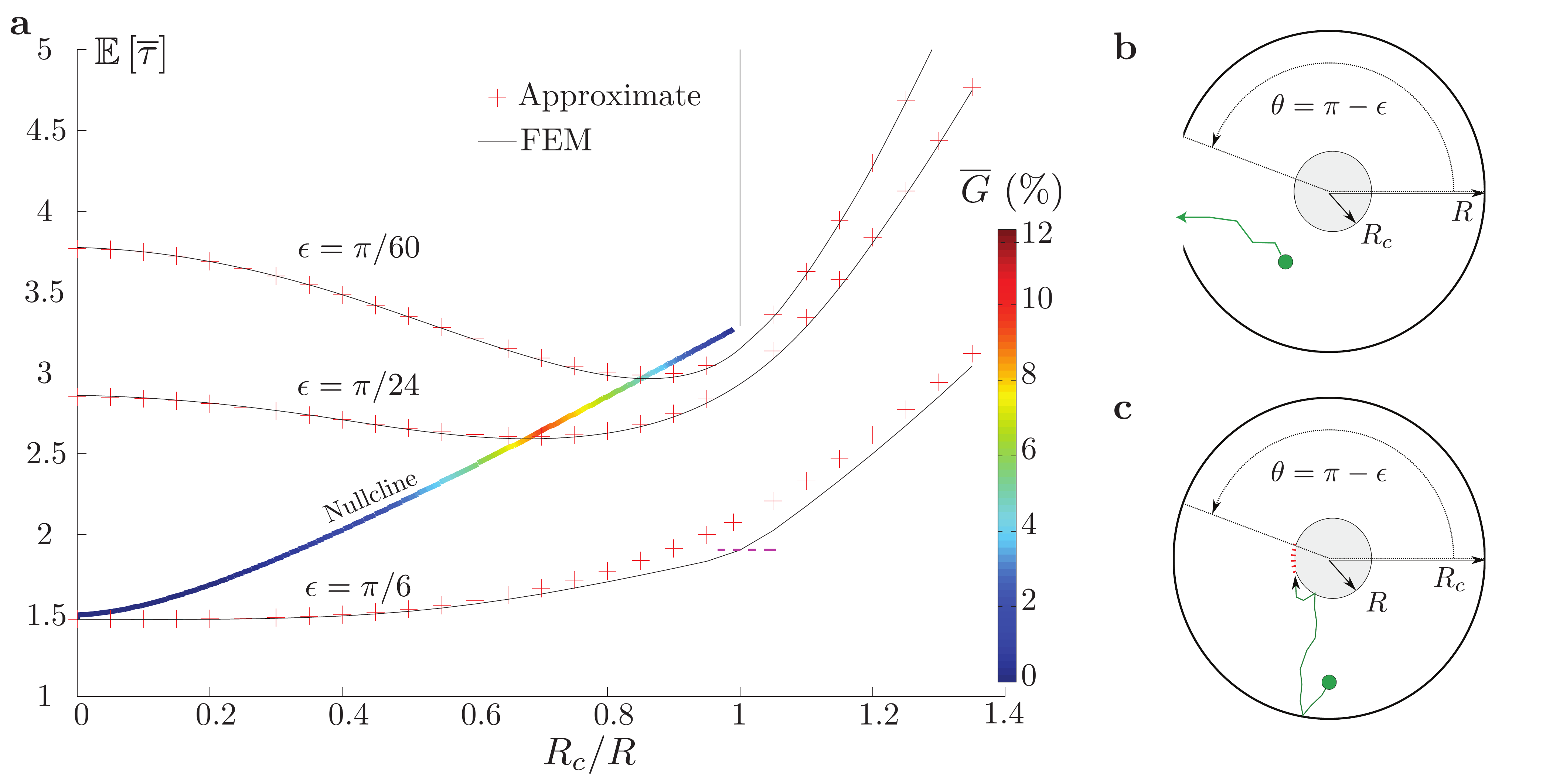}
\caption{
(Color Online) \textbf{(a)} GMFPT to an exit of half-width $\epsilon$
for a Brownian particle confined in an annulus $\mathrm{\Omega}$ of
radii $R_c$ and $R=1$ (illustrated by \textbf{(b)} for $R_c > R$ and
\textbf{(c)} for $R_c < R$).  The approximate solution from
Eq. (\ref{eq:a0_annulus}) (red pluses) is compared to finite element
method simulations (black solid line).  The 1D GMFPT from
Eq. (\ref{eq:mean_1D}), $(\pi - \epsilon)^3/(3\pi)$, is shown by
magenta dotted line for $\epsilon = \pi/6$.  The colored solid line
indicates the loci of the minima of the GMFPT, with the color code
being a function of the gain $\overline{G}$, defined by
Eq. (\ref{eq:gain}).  The gain has a sharp maximum $G
\approx 10 \%$ for $R^{(c)}_c = 0.70$ and $\epsilon = 0.13 \ll 1$.
(\textbf{b,c}) Two annuli with $R_c < 1$ (\textbf{b}) and $R_c > 1$
(\textbf{c}).  The Brownian particle shown by green circle diffuses in
the annulus before crossing the exit of half-width $\epsilon$. }
\label{fig:annulus}
\end{figure}

\subsubsection{Optimization of the GMFPT}

Now we focus on the GMFPT $\overline{\esp{\tau}}$ defined in
Eq. (\ref{def:general_definition_average}).  In
Fig. \ref{fig:annulus}, we present the GMFPT as a function of $R_c$
for different exit sizes $\epsilon$.  Interestingly, for small enough
exit sizes $\epsilon < \epsilon_{c}$, the GMFPT is minimized for a
specific value of the reflecting boundary radius $R^{(c)}_c < 1$.

In the narrow-escape limit $\epsilon \ll 1$, $R^{(c)}_c = 1$ is a
global minimum of the GMFPT: the GMFPT at $R_c =1$ converges to
$\pi^2/3$ while the GMFPT diverges logarithmically with $\epsilon \ll
1$ for any other value of $R_c \neq 1$.  For increasing values of
$\epsilon$, the global minimum of the GMFPT is reached at smaller
values $R^{(c)}_c < 1$.  Eventually, the minimum $R^{(c)}_c = 0$
emerges for exit sizes larger than a threshold: $\epsilon_{c} \simeq
0.51$.

We determine an approximate value for the threshold $\epsilon_{c}$
based on the approximate expression (\ref{eq:averagedmfptrc}) for the
GMFPT.  We first notice that for all $R_c \geq 1$, the GMFPT is a
monotonically increasing function of $R_c$, hence $R^{(c)}_c \leq 1$.
We define $\epsilon_{c}$ as the largest value of $\epsilon$ such that
the GMFPT is a locally decreasing function at $R_c = 0$.  This local
condition is fulfilled if and only if the second derivative of the
GMFPT $\overline{\esp{\tau}}$ is negative at $R_c = 0$, leading to the
following criterion on $\epsilon_{c}$:
\begin{align} \label{eq:criticalepsilon}
\alpha_0(\epsilon_c)  = 4 \alpha_1(\epsilon_c)  - \frac{3}{4} ,
\end{align}
where $\alpha_0$ and $\alpha_1$ are given in
Eqs. (\ref{eq:a0careyphi_pi}) and (\ref{eq:ancareyphi_pi}).  Assuming
$\epsilon_c \ll 1$, Eq. (\ref{eq:criticalepsilon}) can be solved
explicitly to get
\begin{align}
\epsilon_c \approx \exp\left(\frac{8 \ln(2)-13}{16}\right) \approx 0.60,
\end{align}
In turn, the numerical solution of Eq. (\ref{eq:criticalepsilon})
yields $\epsilon_{c} \simeq 0.51$ which is close to the above
estimate.

One may wonder how much time can be gained by setting $R_c$ to the
optimal $R^{(c)}_c$?  We define the gain $\overline{G}$ as
\begin{align} \label{eq:gain}
\overline{G} = \frac{\min\left(\overline{\esp{\tau}}_{R_c = 0} \ , \ \overline{\esp{\tau}}_{R_c =1} \right) 
- \overline{\esp{\tau}}_{R^{(c)}_c}}{\min\left(\overline{\esp{\tau}}_{R_c = 0} \ , \ \overline{\esp{\tau}}_{R_c =1} \right)},
\end{align}
so that $\overline{G}$ lies between $0$ and $1$.  The loci of the
minima of the GMFPT are the set of points $(R^{(c)}_c,
\overline{\esp{\tau}}_{R^{(c)}_c})$ which are shown in
Fig. \ref{fig:annulus} and colored according to the gain
$\overline{G}$.

The gain has a sharp maximum $G \approx 10 \%$ for $R^{(c)}_c \simeq
0.70$ and $\epsilon \simeq 0.13 \ll 1$.  Notice that the optimal gain
is obtained for a value of $\epsilon$ such that the GMFPT at $R_c = 0$
is approximately equal to the GMFPT at $R_c = 1$.  The optimal
$R^{(c)}_c$ results from a trade-off between two competing geometrical
effects: (i) increasing $R_c \ll R$ reduces the accessibility to the
exit for remote particles which have to circumvent the reflecting
boundary at $r = R_c$; and (ii) once a particle is close to the exit,
increasing $R_c$ increases the probability for the particle to cross
the exit.

\subsection{Advection-diffusion with a radial bias} \label{sec:drift}

We consider a diffusive particle confined in a disk of radius $R$
whose motion is biased by a $1/r$ velocity field $\vec{v}(r)$.  The
velocity field $\vec{v}(r)$ is characterized by a dimensionless
parameter $\mu$:
\begin{equation}
\vec{v}(r) = \frac{\mu D}{r^2} \; \vec{r}.
\end{equation} 
Note that $\mu > 0$ corresponds to an outward drift.  Setting units by
$R = 1$ and $D = 1$, the backward diffusion Eq. (\ref{eq:besselfull0})
on the survival probability reads
\begin{align}
\left( \mathrm{\Delta} +\frac{\mu}{r} \partial_r \right)\ S^{(p)}(r,\theta) &= p  \ S^{(p)}(r,\theta) - 1, & 
r \in [0, 1), \quad &  \theta \in [0,2\pi). \label{eq:besselfullRc} 
\end{align}
A separation of variables method provides the set $f^{(p)}_n \cos(n
\theta)$ of solutions for Eq. (\ref{eq:besselfullRc}), see Table
\ref{tab:ptable}.

\begin{figure}[t]
\centering
\includegraphics[height=7.5cm]{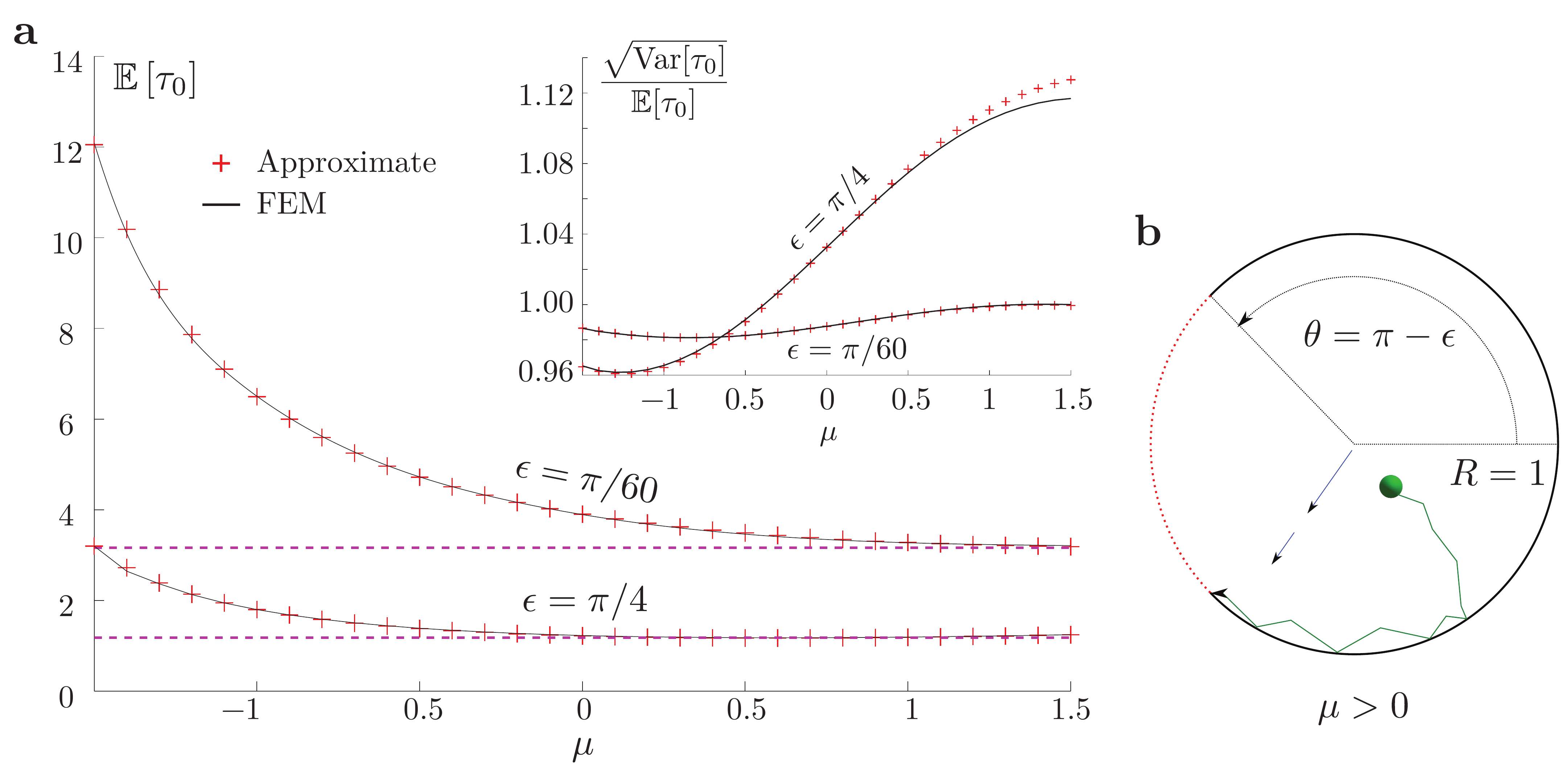}
\caption{
(Color Online) (\textbf{a}) The MFPT $\esp{\tau_0}$ to an exit of
half-width $\epsilon = \pi/4$ for Brownian particles whose diffusive
motion is biased by a $1/r$ velocity field $\vec{v}(r) = \mu D
\vec{r}/r^2$.  Particles are started at $r = 0$ inside the unit disk
[as sketched in (\textbf{b})].  The analytical approximation in
Eq. (\ref{eq:a0_fulldisk_drift}) (red pluses) is compared to a finite
element method (denoted FEM, black solid line).  The 1D GMFPT from
Eq. (\ref{eq:mean_1D}), i.e., $(\pi - \epsilon)^3/(3 \pi)$, is shown
by horizontal magenta dashed lines.  The ratio of the standard
deviation to the MFPT is represented in the inset.  Note that the
smaller the $\epsilon$, the closer this ratio is to $1$.  \textbf{(b)}
A Brownian particle (shown by green circle) is advected by a radial
flow field $\vec{v}(r) = \mu D\; \vec{r}/r^2$, with $\mu >0$
corresponding to an outward drift (blue arrows).  The particle is
reflected by the boundary at $r = 1$ before crossing the exit (shown
by red dashed line) of half width $\epsilon = \pi/4$. }
\label{fig:drift}
\end{figure}

The approximate scheme of Sec. \ref{sec:approximate} leads to an
explicit expression for the MFPT.  We substitute $\mathcal{C}^{(0)}$
by $\mathcal{C}_a^{(0)}$ (defined in Eq. (\ref{eq:substitution})) into
Eqs. (\ref{eq:a0exact}) and (\ref{eq:anexact}) and use the expressions
of Table \ref{tab:ptable} to get the following approximation for the
Fourier coefficients of the MFPT:
\begin{align} \label{eq:a0_fulldisk_drift}
a_{0}^{(0)} &\approx \frac{2}{2+\mu}\left\lbrace \alpha_0 + \sum^{\infty}_{k = 1} 2k \left[-1 + \frac{\mu}{2k}  + 
\sqrt{1+\left(\frac{\mu}{2k} \right)^2} \right] \alpha^2_k \right\rbrace   , \\
a_{n}^{(0)} &\approx  \frac{2}{2+\mu}~ \frac{\alpha_{n} }{\sqrt{1+\left(\frac{\mu}{2n} \right)^2} -\frac{\mu}{2n}}, 
\qquad n \geq 1. \label{eq:an_fulldisk}
\end{align}
where $\alpha_n$ are defined by Eqs. (\ref{eq:ancarey}).  The MFPT can
be computed from Eq. (\ref{eq:mfpt_general}). The GMFPT, defined in
Eq. (\ref{def:general_definition_average}), is
\begin{align} \label{eq:averagedmfptmu}
\overline{\esp{\tau}} \approx  \frac{1}{2+\mu}\left\lbrace \alpha_0 + \sum^{\infty}_{k = 1} 2k \left[-1 + \frac{\mu}{2k} 
 + \sqrt{1+\left(\frac{\mu}{2k} \right)^2} \right] \alpha^2_k \right\rbrace
+ \frac{1}{4(2+\mu)}.
\end{align}
In Fig. \ref{fig:drift}, we compare the approximate MFPT
$\esp{\tau_0}$ (for a particle started at $r=0$) to the result
obtained by a finite element method.  The MFPT diverges for all $\mu
\leq -2$: the inward drift strongly confines particles at $r = 0$.  In
the limit $\mu \gg 1$, the MFPT converges to the MFPT of a 1D process
given by Eq. (\ref{eq:mean_1D}), up to a small $\mathcal{O}(\epsilon)$
correction.  The MFPT is a monotonically decreasing functions of
$\mu$, as illustrated in Fig. \ref{fig:drift}(a), and there is no
optimal drift which minimizes the MFPT.

\subsection{Rectangles} \label{sec:rectangle}

\begin{figure}[t]
\centering
\includegraphics[scale=0.37]{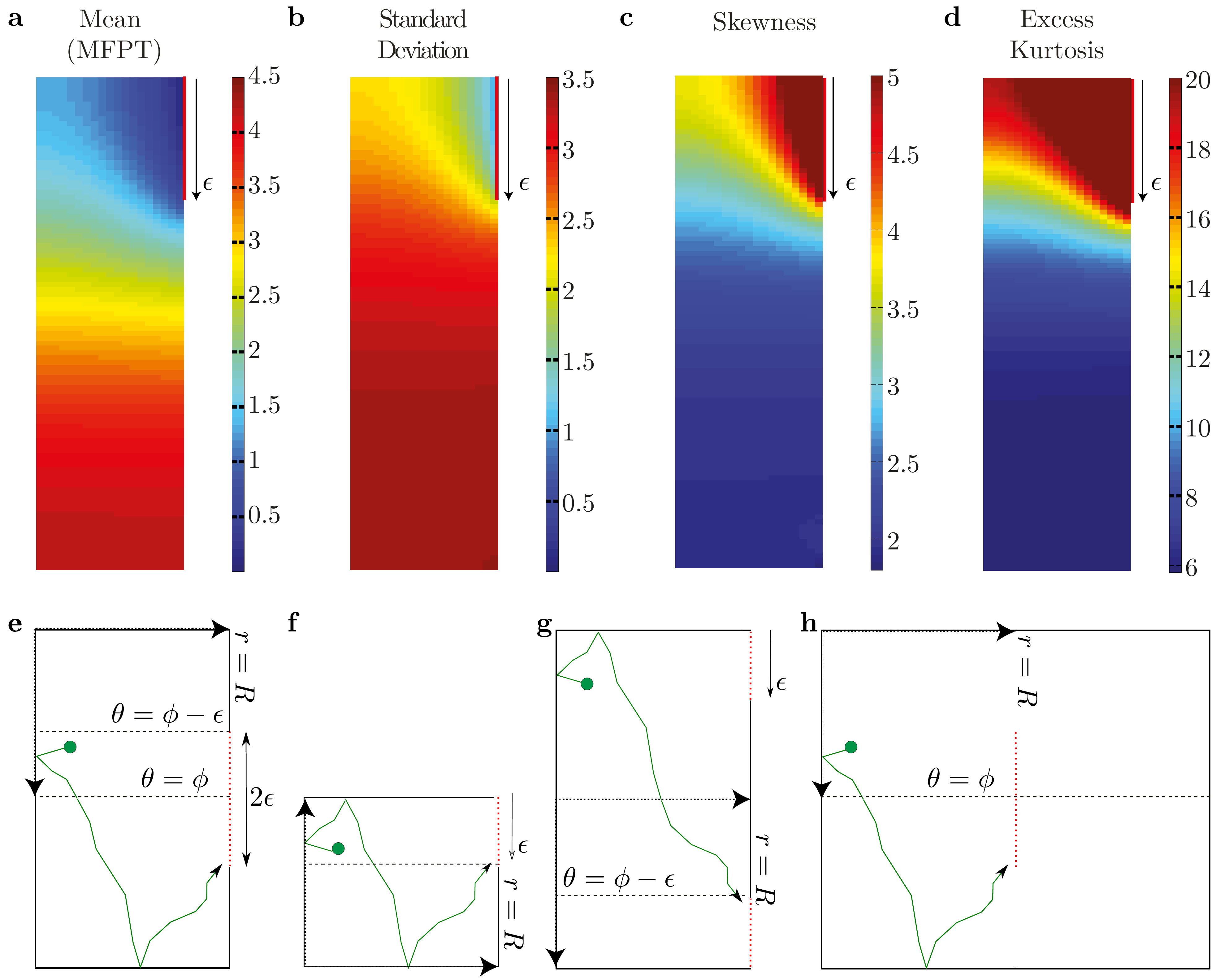} 
\caption{
(Color Online) \textbf{Upper panel}: The cumulants of the exit time as
functions of the starting position of the Brownian particle within the
rectangle $\mathrm{\Omega} = \left[ 0, R \right] \times \left[ 0,
\phi\right]$ with $R = 1$ and $\phi = \pi$.  The exit (shown by red
line and an arrow) is a linear segment of total width $\epsilon =
\pi/4$ on the edge of length $\pi$.  The cumulants computed through
the approximate resolution of Eqs. (\ref{eq:a0_rectangle}) and
(\ref{eq:an_rectangle}): (\textbf{a}) MFPT; (\textbf{b}) standard
deviation; (\textbf{c}) skewness; and (\textbf{d}) excess kurtosis.
\textbf{Lower panel}: The distribution of the FPT is identical for
the four cases: (\textbf{e}) rectangle $\mathrm{\Omega} = [0,R] \times
[0, 2\phi]$ with reflecting walls, pierced by a centered opening of
width $2 \epsilon$; (\textbf{f}) rectangle $\mathrm{\Omega} = [0, R]
\times [0,\phi]$ with reflecting walls, pierced by an opening of width
$\epsilon$ located in a corner; (\textbf{g}) rectangle
$\mathrm{\Omega} = [0, R] \times [0, 2\phi]$ with reflecting walls,
pierced by two cornered openings, each of width $\epsilon$; and
(\textbf{h}) rectangle $\mathrm{\Omega} = [0, 2R] \times [0, 2\phi]$
with reflecting walls and a centered linear absorbing region (vertical
red dashed line) of total width $2 \epsilon$.  Green circle represents
the position of a particle in these rectangles.}
\label{fig:moment_rectangle}
\end{figure}

We consider the confining domain $\mathrm{\Omega}$ to be a rectangle
$\mathrm{\Omega} = [0, R] \times [0, \phi ]$ with reflecting edges at
$r = 0$, $\theta = 0$ and $\theta = \phi$.  The boundary at $r = R$ is
reflecting except for an absorbing segment of length $\epsilon$ at the
corner, as illustrated in Fig. \ref{fig:moment_rectangle}(b).  Setting
units by $\phi = \pi$ and $D = 1$, the Helmholtz equation on the
survival probability reads \cite{Redner:2001a,Gardiner:2004}
\begin{align} \label{eq:sinrectangle}
\left( \pdds{}{r}  + \pdds{}{\theta} \right) S^{(p)}(r,\theta) = p \ S^{(p)}(r,\theta) - 1 
\qquad & (r, \theta) \in \mathrm{\Omega}.
\end{align}

Although our approach allows one to get the whole FPT distribution, we
focus on obtaining an explicit expression for the MFPT.  We substitute
$\mathcal{C}^{(0)}$ by $\mathcal{C}_a^{(0)}$ (defined in
Eq. (\ref{eq:substitution})) in Eqs. (\ref{eq:a0exact}) and
(\ref{eq:anexact}) and use the expressions of Table \ref{tab:ptable}
to get the following approximation for the Fourier coefficients of the
MFPT:
\begin{align} \label{eq:a0_rectangle}
a_{0}^{(0)} &\approx  2 R \left\lbrace \alpha_0 + \sum^{\infty}_{k = 1} \left[ -1+\coth(k R)  \right] 4k \alpha^2_k \right\rbrace
, \\
a_{n}^{(0)} &\approx \frac{2 R}{\coth(nR)} \alpha_n , \qquad n \geq 1, \label{eq:an_rectangle}
\end{align}
where $\alpha_n$ are defined Eqs. (\ref{eq:ancarey}).  Using
Eq. (\ref{eq:a0_rectangle}) and (\ref{eq:an_rectangle}), the MFPT can
be computed from Eq. (\ref{eq:mfpt_general}).  Figure
\ref{fig:moment_rectangle} shows the approximate MFPT as a function of
the initial position $(r, \theta) \in \mathrm{\Omega}$ for $R = 1$ and
$\phi = \pi$.  We also compute the standard deviation, skewness, and
excess kurtosis through the approximate resolution scheme of
Eqs. (\ref{eq:a0_rectangle}) and (\ref{eq:an_rectangle}).  The error
of the approximate solution to the numerical and exact resolution
scheme is below $2 \%$ (the error is maximal close to the edges of the
exit).

The GMFPT, defined in Eq. (\ref{def:general_definition_average}), is
\begin{align}
\overline{\esp{\tau}} \approx  R \left\lbrace \alpha_0 + 
\sum^{\infty}_{k = 1} \left[ -1+\coth(k R)  \right] 4k \alpha^2_k \right\rbrace + \frac{R^2}{3} .
\end{align}
In the limit $R \ll \phi = \pi$, the GMFPT converges to the GMFPT of a
1D process given by Eq. (\ref{eq:mean_1D}), up to small
$\mathcal{O}(\epsilon)$ correction.  In the opposite limit $R \gg 1$,
the relation
\begin{align}
 -1+\coth(k R) = -1 + \frac{1 + \exp(-2 k R)}{1 - \exp(-2 k R)}= 2 \beta^2 + \mathcal{O}(\beta^4)
\end{align}
with $\beta = \exp\left(- R \right) $, leads to 
\begin{align} \label{eq:a0_rectangle_greatRr}
a_{0}^{(0)} &\approx  4 R\left[ \ln\left( \frac{2}{\epsilon} \right) + 2\beta^2 \right] + \mathcal{O}(\beta^4)
\end{align}
in the small $\epsilon \ll 1$ limit.  The latter expression can be
identified with the result presented in \cite{Singer:2006c} (p. 496),
which is shown to be exact up to a $\mathcal{O}(\epsilon)$ term.

\subsection{Analogy to heat transfer problems} \label{sec:heat}

We show that the resolution scheme of Sec. \ref{sec:resolution}
provides the evolution of temperature in a room $\mathrm{\Omega}$ with
adiabatic walls which include a centered window of width $2\epsilon$.
In fact, the evolution of temperature $\tilde{T}^{(t)}(r,
\theta)$ is governed by the heat equation \cite{Carslaw1959,Crank1975}
\begin{align} \label{eq:difftemp}
\frac{\partial \tilde{T}^{(t)}(r, \theta)}{\partial t}= D \mathrm{\Delta} \tilde{T}^{(t)}(r, \theta).
\end{align}
The temperature in the room $\mathrm{\Omega}$ is supposed to be
homogeneous before opening the window at time $t= 0$:
$\tilde{T}^{(0)}(r, \theta) = T_1$.  Since that moment and for all $t
>0$, the boundary condition at the window is $\tilde{T}^{(t)}(r,
\theta) = T_0$, where $T_0$ is an exterior temperature.  Along the
adiabatic walls the temperature $\tilde{T}^{(t)}(r, \theta)$ satisfies
the reflecting boundary condition.  The function
\begin{align}
\tilde{w}^{(t)}(r, \theta) = \frac{\tilde{T}^{(t)}(r, \theta) - T_0}{T_1 - T_0}.
\end{align}
satisfies (i) the heat Eq. (\ref{eq:difftemp}), (ii) Dirichlet
boundary condition at the window ($\tilde{w}^{(t)}(r, \theta) = 0$ for
all $t > 0$), (iii) Neumann boundary condition along the adiabatic
walls, and (iv) the initial condition $\tilde{w}^{(0)}(r, \theta) =
1$.  Since the survival probability satisfies the same equations, one
concludes that $\tilde{w}^{(t)}(r, \theta) = \tilde{S}^{(t)}(r,
\theta)$.


\subsection{Analogy to microchannel flows} \label{sec:microchannelflow}

\begin{figure}[h!]
\centering
\includegraphics[height=4.8cm]{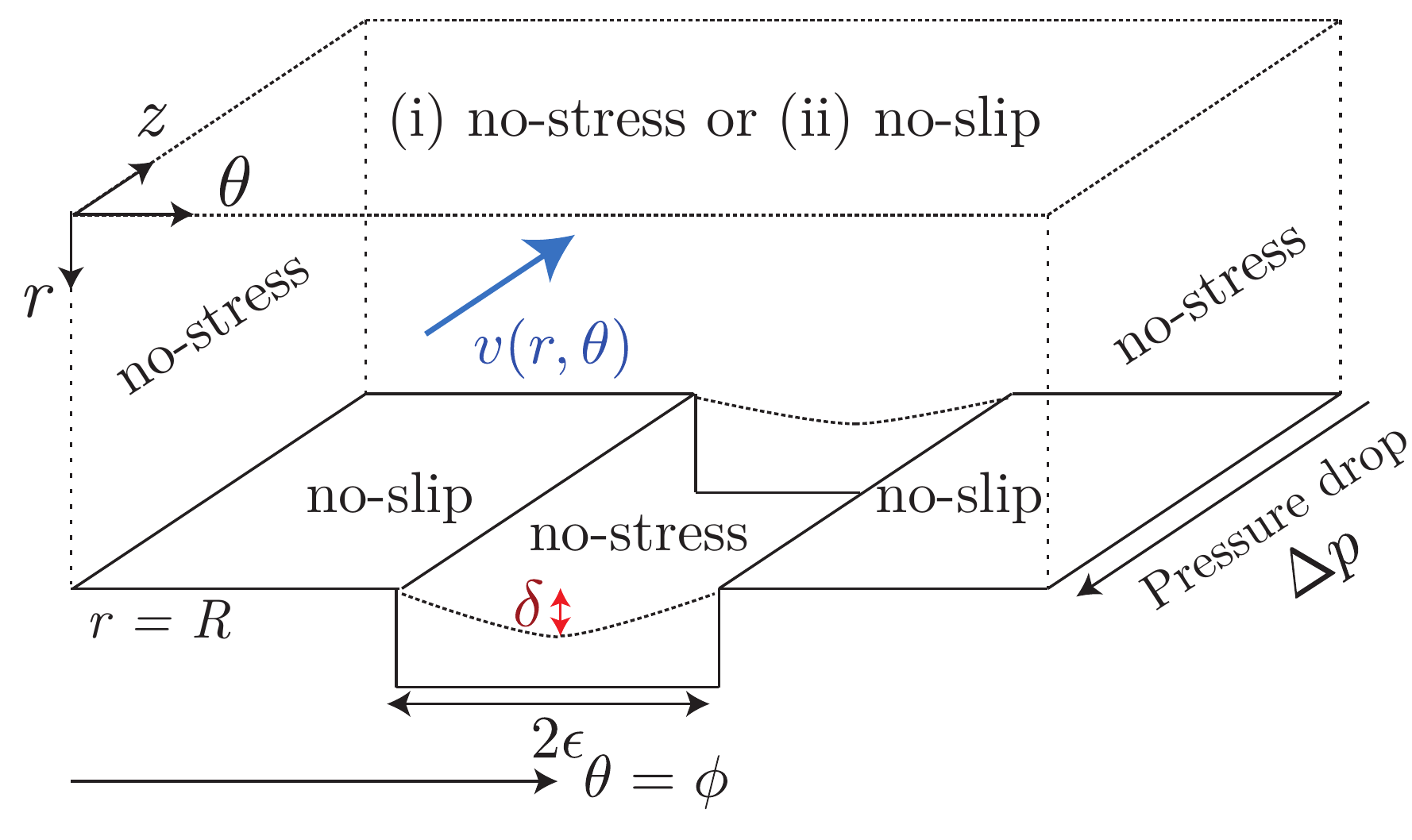}
\caption{
(Color Online) Scheme of the microchannel flow problem, in which the
floor of a channel of depth $R$ contains a large number of regularly
spaced grooves of width $2\epsilon$ parallel to the flow direction
$(0z)$.  This structure can be modeled by the periodic repetition of a
fundamental cell of width $2\phi$, resulting in a no-shear condition
$\theta = 0$ and $\theta = 2\phi$. The shear stress is assumed to be
zero along the free surfaces within the groove at $r =R$ (the free
surface lies above trapped gas phase).  In turn, non-slip boundary
condition is imposed on the remaining part of the groove.  The top
surface at $r = 0$ can be assumed to be: (i) a free surface along
which the shear stress is assumed to be zero (ii) a no-slip surface
(i.e. the case considered in Ref. \cite{Sbragaglia2007}). The problem
consists in determining the stationary velocity profile
$v^{(\infty)}(r, \theta)$ for an incompressible Newtonian fluid at low
Reynolds numbers and under constant pressure drop.}
\label{fig:11}
\end{figure}

Large pressure drops are necessary to cause liquid flow in
microchannels due to viscous dissipation at the boundary (no-slip
condition).  In order to increase the flow rate (at a given pressure
drop), one can introduce ultra-hydrophobic grooves so that the layer
of gas trapped within the grooves would act as an air-cushion for the
fluid flow \cite{Joseph2006,Cottin-Bizonne2004}.  We consider an array
of ultra-hydrophobic grooves aligned in the direction of the pressure
drop $z$ obtained by a periodic repetition of a fundamental cell of
width $\theta = 2 \phi$.  The floor of the microchannel is at the
depth $r = R$ (see Fig. \ref{fig:11}).  The top surface at $r = 0$ can
be assumed to be either (i) a free surface such that the shear stress
is equal to zero, or (ii) a no-slip surface (as considered in
Ref. \cite{Sbragaglia2007}).

In the case of a no-slip condition at $r = R$ (case (ii)) and in the
limit $R \gg \phi$, an exact solution of the stationary flow was found
in terms of the set $(\alpha_n)$ defined in
Eqs. (\ref{eq:a0careyphi_pi}) and (\ref{eq:ancareyphi_pi})
\cite{Sbragaglia2007}.  In this section we show how our method can be
adapted to provide: (i) an approximate solution for the flow which is
accurate for any value of $R$, and (ii) an exact resolution scheme as
well as an approximate explicit expression for a time-dependent
problem, i.e., the evolution of the flow from a given radial profile
at $t = 0$ to the steady state profile at $t = \infty$.

The flow is assumed to be (i) Newtonian and incompressible, (ii) at
zero Reynolds number, (iii) in the absence of external force
(e.g. gravitational force), and (iv) under a constant pressure
gradient $\pd{p}{z} = q$.  Under these assumptions the Navier-Stokes
equation on the velocity profile $\tilde{v}^{(t)}(r,\theta)$ reads as
\begin{align} \label{eq:navierstockes}
\rho \pd{\tilde{v}^{(t)}(r, \theta)}{t} = \mu \mathrm{\Delta} \tilde{v}^{(t)}(r, \theta) +  q, 
\qquad (r, \theta) \in [0, \, R] \times [0 , 2\phi] ,
\end{align}
where $\rho$ is the mass density of the fluid and $\mu$ its viscosity.
In dimensionless variables $\theta \leftarrow \pi \theta/\phi$, $R
\leftarrow \pi R/\phi$, $\tilde{v}^{(t)} \leftarrow (\pi/\phi)^2
(\mu/q) \tilde{v}^{(t)}$ and $t \leftarrow (\mu \phi^2 t)/(\rho
\pi^2)$, Eq. (\ref{eq:navierstockes}) becomes
\begin{align} \label{eq:navierstockesadimension}
\pd{\tilde{v}^{(t)}(r, \theta)}{t}  = \mathrm{\Delta} \tilde{v}^{(t)}(r, \theta) + 1, \qquad (r, \theta) \in [0, \, R] \times [0 , \, 2\pi]
\end{align}
In the stationary regime ($t = \infty$),
Eq. (\ref{eq:navierstockesadimension}) reads
\begin{align} \label{eq:flowstationnary}
\mathrm{\Delta} \tilde{v}^{(\infty)}(r, \theta) = - 1, \qquad (r, \theta) \in [0, \, R] \times [0 , \, 2\pi ].
\end{align}
The latter equation on the stationary flow $\tilde{v}^{(\infty)}(r,
\theta)$ can be identified with the equation on the MFPT
(e.g. Eq. (\ref{eq:besselfull0}) at $p =0$).  The Laplace transform of
Eq. (\ref{eq:navierstockesadimension}) is
\begin{align} 
\label{eq:navierstockesadimension_LP}
\mathrm{\Delta} \, v^{(p)}(r,\theta)  &= p  \ v^{(p)}(r,\theta) - \tilde{v}^{(0)}(r) - \frac{1}{p} ,
\end{align}
where $\tilde{v}^{(0)}(r)$ is the initial velocity profile at $t = 0$,
which is assumed to be independent of $\theta$.  Note that the
long-time flow profile $\tilde{v}^{(\infty)}(r,\theta)$ can be deduced
from $v^{(p)}(r,\theta)$ through the relation:
\begin{align} \label{eq:limitvinf}
\lim\limits_{p \rightarrow 0} \ p \, v^{(p)}(r,\theta) = \tilde{v}^{(\infty)}(r,\theta).
\end{align}
Eq. (\ref{eq:navierstockesadimension_LP}) is completed by the
following boundary conditions.  The shear stress is assumed to be zero
along the free surfaces, i.e., at $\theta = 0$, $\theta = 2\pi$, $r =
0$, and within the groove at $r =R$ (the free surface lies above the
gas trapped within the groove).  We consider the case $\delta = 0$,
where $\delta$ is the maximum penetration of the free surface into the
groove.  This approximation is justified because the surface of the
groove is hydrophobic.  At the bottom surface $r = R$, the velocity
field satisfies the mixed boundary conditions:
\begin{itemize}
\item 
non-slip conditions along the hydrophobic surface: $v^{(p)}(r,\theta)
= 0$ for all $\theta \in [\pi - \epsilon, \pi + \epsilon]$ (similar to
Eq. (\ref{eq:bc1u})),

\item 
no-shear conditions along the free surface: $\left[\pa_r
v^{(p)}(r,\theta)\right]_{r=R}= 0$ for all $\theta \in [0, \pi -
\epsilon) \cup (\pi + \epsilon, 2\pi]$ (similar to
Eq. (\ref{eq:bc2u})).
\end{itemize}
Similarly to Eq. (\ref{eq:auxillary}), we define the auxiliary
function
\begin{equation} \label{eq:auxillary_micro}
u^{(p)}(r,\theta) \equiv  v^{(p)}(r,\theta) - v^{(p)}_{\pi}(r),
\end{equation}
where $v^{(p)}_{\pi}(r)$ is the rotation invariant solution of
Eq. (\ref{eq:navierstockesadimension_LP}) satisfying $v^{(p)}_{\pi}(1)
= 0$, and either $\pa_{r} S^{(p)}_{\pi}(r) = 0$ at $r = 0$ for a free
surface (i), or $v^{(p)}_{\pi}(1)= 0$ at $r = 0$ for a no-slip surface
(ii).

The Fourier expansion of the function $u^{(p)}(r,\theta)$ according to
Eq. (\ref{eq:FourierBessel}) defines the Fourier coefficients
$a^{(p)}_n$.  In the case of free surface, functions $f^{(p)}_n$ are
given in Table \ref{tab:ptable}.  In the case of a no-slip surface,
functions $f^{(p)}_n$ read
\begin{align}
f^{(p)}_n(r) = \frac{\sinh(\sqrt{p + n^2} \ r)}{\sinh(\sqrt{p + n^2} R)}, \qquad n \geq 0.
\end{align}
The Fourier coefficients $a^{(p)}_n$ are shown to satisfy
Eqs. (\ref{eq:a1Sneddon}) and (\ref{eq:a2Sneddon}).  One can therefore
apply the resolution scheme presented in Sec. \ref{sec:resolution} to
derive both an exact and an approximate expression for the Laplace
transform $v^{(p)}(r,\theta)$ of the flow velocity.  An approximate
expression for stationary velocity profile $\tilde{v}^{(\infty)}(r,
\theta)$ is then deduced from Eq. (\ref{eq:limitvinf}).

\section*{Conclusion}

We studied the Helmholtz equation with mixed boundary conditions on
spherically symmetric two-dimensional domains (disks, angular sectors,
annuli).  This classical boundary value problem describes how
diffusive particles exit from a domain through an opening on the
reflecting boundary.  The Dirichlet boundary condition on the opening
is mixed with Neumann boundary condition on the remaining part of the
boundary that presents the major challenge in the resolution of this
problem.  For this reason, most previous studies were focused on the
asymptotic analysis for small exits.  In order to overcome this
limitation, we developed a new approach, in which the problem is
reduced to a set of linear equations on the Fourier coefficients of
the survival probability.  We provide then two resolution schemes
which are applicable for arbitrary exit size.  The first scheme is
exact but it relies on a numerical solution of linear equations and
requires thus a matrix inversion.  In turn, the second scheme is
explicit (without matrix inversion) but approximate.  As a result, we
managed to derive the whole distribution of first
passage times and their moments for the escape problem with arbitrary
exit size.  The approximate solution was shown to be accurate over the
whole range of times.  Both analytical solutions have been
successfully verified by extensive numerical simulations, through both
a finite element method resolution of the original boundary value
problem, and by Monte Carlo simulations.

Using this method, we analyzed the behavior of the FPT probability
density for various initial positions.  When the initial position is
far from the exit, the FPT probability density was shown to be
accurately approximated by an exponential distribution.  In this
situation, the whole distribution of FPTs is essentially determined by
the MFPT for which we derived exact explicit relations.

The developed method is also applied to rectangular domains and to
biased diffusion with a radial drift within a disk.
%
Since the Helmholtz equation with mixed boundary conditions is also
encountered in microfluidics \cite{Sbragaglia2007}, heat propagation
\cite{Carslaw1959,Crank1975}, quantum billiards \cite{Castro2005,Grebenkova},
and acoustics \cite{Temkin2001}, the developed method can find
numerous applications beyond first passage processes.

\begin{acknowledgments}
O.B. is supported by the ERC Starting Grant No. FPTOpt-277998.
D.G. is supported by an ANR project ``INADILIC''.

The final publication is available at Springer via:
\begin{center}
http://link.springer.com/article/10.1007{\%}2Fs10955-014-1116-6.
\end{center}
\end{acknowledgments}

\appendix
\section{Simplification of $\alpha_{n}$ and $M_{nm}$} \label{eq:a0an}

\subsection{Simplified expressions for $\alpha_{0}$}

Herewith we prove the following identity for all $0 \leq t < \pi$
\begin{align}
\frac{\sqrt{2}}{\pi}  \int_{0}^{t} \mathrm{d}x \frac{x \sin(x/2)}{\sqrt{\cos x - \cos t }}
=    - 2 \ln \left(\cos \left( \frac{t}{2}\right) \right) \label{eq:a1}.
\end{align} 
We proceed by a change of variable $z = \cos x$ in the left-hand side
term of Eq. (\ref{eq:a1}) and we denote $T = \cos (t)$:
\begin{align}
\int_{0}^{t} \mathrm{d}x \frac{x \sin(x/2)}{\sqrt{\cos x - \cos t }}
=   \int_{T}^{1} \mathrm{d}z \left[\frac{\arccos z}{\sqrt{2\left(1+z\right)}} \right] \frac{1}{\sqrt{z - T }}  \label{eq:kern1}.
\end{align} 
We write the right-hand side of Eq. (\ref{eq:a1}) in the form
\begin{align}
- 2 \log \left(\cos\left( \frac{t}{2} \right) \right) =  \log \left( \frac{2}{1+ T} \right).
\end{align}
From Ref. \cite{Sneddon1966}, the Abel's equation
\begin{align}
\int^{1}_{T} \frac{y(z) \mathrm{d}z}{\sqrt{ z - T}} = \frac{\pi}{\sqrt{2}} \log \left( \frac{2}{1+ T} \right)
\end{align}
has an unique solution for all $ -1 < X < 1$
\begin{align} \label{eq:kern2}
y(z) &= \frac{1}{\pi} \frac{\pi}{\sqrt{2}} \int^{z}_{1} \frac{\mathrm{d}u}{\sqrt{u -z}(1+z)} = \frac{\arccos{z}}{\sqrt{2 (1 + z)}}.
\end{align}
Identification of the kernels of Eqs. (\ref{eq:kern1}) and
(\ref{eq:kern2}) proves the identity (\ref{eq:a1}).

The expression for $\alpha_{0}$ from Eq. (\ref{eq:a0carey}) is found
by setting $t = \pi - \epsilon$ in Eq. (\ref{eq:kern1}).  Note that
the obtained expression for $\alpha_{0}$ from Eq. (\ref{eq:a0carey})
could also be deduced from the expression of the MFPT from
Ref. \cite{Caginalp2012}.

\subsection{Simplified expressions for $\alpha_{n}, n \geq 1$}

The solution of Eqs. (\ref{eq:bc1p0}), (\ref{eq:bc2p0}) is given in
\cite{Singer:2006b} in the form:
\begin{align}
\alpha_{n} &= \frac{1}{\sqrt{2} \pi} \int_{0}^{\pi - \epsilon}\! \! \!  
\mathrm{d}t \left(  \pd{}{t} \int_{0}^{t} \mathrm{d}x  \frac{x \sin(x/2)}{\sqrt{\cos x - \cos t }} \right) 
\bigl[ P_n(\cos t) + P_{n-1}(\cos t) \bigr], \qquad n \geq 1.
\end{align} 
Using the identity (\ref{eq:a1}), we show that
\begin{align}
\alpha_{n} &= \frac{1}{2} \int_{0}^{\pi - \epsilon} \! \! \! \mathrm{d}t \tan\left(\frac{t}{2}\right) 
\bigl[ P_n(\cos t) + P_{n-1}(\cos t) \bigr], \qquad n \geq 1.
\end{align} 
After the change of variable $ u = \cos t$, the latter identity leads
to
\begin{align} 
\alpha_{n} &= \frac{1}{2}\int^{1}_{-\cos \epsilon}  \frac{\mathrm{d}u}{1+u} \bigl[ P_n(u) + P_{n-1}(u)\bigr] , \qquad n \geq 1.
\end{align} 
We now use the identity
\begin{align}  \label{eq:myfavoritederivative}
\pd{}{x} \left( \frac{P_{m}(x)-P_{m-1}(x)}{m}\right)_{\lvert x = X} = \frac{P_{m}(X) + P_{m-1}(X)}{1 + X}, \qquad m \geq 1,
\end{align}
which is valid for all $X \in \left[ - 1 , 1 \right]$, to obtain the
announced result:
\begin{align}
\alpha_{n} &= \frac{(-1)^{n-1}}{2 n} \bigl[ P_{n}(\cos \epsilon) + P_{n-1}(\cos \epsilon) \bigr]  , \qquad n \geq 1.
\end{align}

\subsection{Simplified expression for $M_{nm}$} \label{sec:Mmatrix}

The expression for $M_{nm}$ from Eq. (\ref{eq:munsimplified}) can be
simplified using Mehler's integral representation (\ref{eq:mehler}):
\begin{align} \label{eq:mnappendix}
M_{nm} &= \frac{1}{2}  \int_{0}^{\pi - \epsilon} \! \! \! \mathrm{d}t \left\lbrace  \pd{}{t} 
\left[P_{m}(\cos t)-P_{m-1}(\cos t)\right] \right\rbrace \bigl[ P_n(\cos t) + P_{n-1}(\cos t) \bigr],  \qquad n \geq 1, \quad  m \geq 1.
\end{align}
The identity Eq. (\ref{eq:myfavoritederivative}) then leads to the
announced expression (\ref{eq:Mepsilon}).  Notice that the following
matrix
\begin{align}
S_{nm} = \sqrt{\frac{n}{m}}M_{nm},  \qquad n \geq 1, \quad  m \geq 1,
\end{align}
is symmetric.  The eigenvalues of $S_{nm}$ are real, and so are the
eigenvalues of $M$.  In Sec. \ref{sec:app_sum} we show that the
coefficients $\alpha_n$ are given by an eigenvector of the
infinite-dimensional matrix $M$.

\subsection{Perturbative expansion of $M_{nm}$} \label{sec:app_perturbatif}

We first derive an alternative identity to
Eq. (\ref{eq:myfavoritederivative}).  Let us define
\begin{align}
A_n(x) &\equiv  \pd{}{x} (P_{n}(x) - P_{n-1}(x)), \qquad n \geq 1.
\end{align}
Using the recurrence formulas for Legendre polynomials, we obtain
\begin{align}
A_n(x)  &= (P'_{n}(x) -P'_{n-2}(x))- (P'_{n-1}(x) -P'_{n-2}(x)) 
 = (2n-1)P_{n-1}(x) - A_{n-1}(x)
 = \sum^{n}_{k=1} (-1)^{n-k} (2k-1)P_{k-1}(x) , \label{eq:Anlegendre}
\end{align}
where we used $P_{0}(x) = 1$, $P_{1}(x) = x$, and $A_1(x) = 1 = (2-1)
P_{0}(x)$.

Combining Eq. (\ref{eq:mnappendix}) and the identity
(\ref{eq:Anlegendre}) we obtain the announced result:
\begin{align}
M_{nm} &= \int_{0}^{\pi - \epsilon} \! \mathrm{d}t \  \frac{\sin t}{2} A_m (\cos t) \bigl[ P_n(\cos t) + P_{n-1}(\cos t) \bigr], \nonumber
\\ &= \sum^{m}_{k=1} (-1)^{n-k} (2k - 1) \left( K_{k-1,n-1} + K_{k-1,n}   \right) ,
\label{eq:Msum}
\end{align}
where the coefficients $K_{k,n}$ are defined by
\begin{align}
K_{k,n} \equiv \int_{0}^{\pi - \epsilon} \! \! \! \mathrm{d}t \ P_k(\cos t) P_n(\cos t) \frac{\sin t}{2} .
\end{align}
In the leading order in $\epsilon \ll 1$, Eq. (\ref{eq:Msum}) reads
\begin{align}
M_{nm} &= \sum^{m}_{k=1} (-1)^{n-k} (2k - 1) \left( 
\frac{\delta_{k-1,n-1}}{2n-1}  + \frac{\delta_{k-1,n}}{2n+1}   \right)  + \mathcal{O}(\epsilon) .
\end{align}
If $m < n$, it is straightforward to show that $M_{nm} =
\mathcal{O}(\epsilon)$.  If $m > n$, $M_{nm} = \mathcal{O}(\epsilon)$
as successive terms with $k = n$ and $ k = n +1$ cancel each other.
The matrix $M_{nm}$ ($n,m \geq 1$) is thus diagonal at the first order
in $\epsilon$:
\begin{align}
M_{nm} =  \delta_{nm} + \mathcal{O}(\epsilon).
\end{align}

In order to get the next term in the series expansion in $\epsilon \ll
1$, we write
\begin{align} \label{eq:order4}
M_{nm} - \delta_{nm} =  - \frac{m}{2} \int^{-\cos \epsilon}_{-1} \frac{1}{1+x} \bigl[ P_m(x)+P_{m-1}(x) \bigr]
\bigl[ P_n(x)+P_{n-1}(x) \bigr] \mathrm{d}x,  \qquad n \geq 1, \quad  m \geq 1.
\end{align}
We now focus on the term in the right-hand side of
Eq. (\ref{eq:order4}).  In the vicinity of $x = -1$, the integrand of
Eq. (\ref{eq:order4}) expands into
\begin{align} \label{eq:lambdanm}
\frac{1}{1+x} \bigl[ P_m(x)+P_{m-1}(x) \bigr] \bigl[ P_n(x)+P_{n-1}(x) \bigr]
= \frac{nm (-1)^{n+m}}{8} (1 +x)+ \mathcal{O}(1+x).
\end{align}
Notice that
\begin{align} \label{eq:secdonc}
\int^{-\cos \epsilon}_{-1}  (1 + x) \mathrm{d}x = \frac{\epsilon^4}{8} + \mathcal{O}(\epsilon^5). 
\end{align}
Substituting Eqs. (\ref{eq:lambdanm}) and (\ref{eq:secdonc}) into
Eq. (\ref{eq:order4}) leads to
\begin{align}
M_{nm}  = \delta_{mn} + \frac{nm^2 (-1)^{n+m}}{8} \epsilon^4 +  \mathcal{O}(\epsilon^5). 
\end{align}

\subsection{Summation identities} \label{sec:app_sum}

Using the identities for sums of Legendre polynomials from
Ref. \cite{Duffy2007}, we derive the following equation
\begin{align} \label{eq:fund}
S \equiv \sum^{\infty}_{m = 1} (-1)^{m-1} \bigl[ P_{m}(\cos x) + P_{m-1}(\cos x) \bigr] 
\bigl[ P_{m}(\cos \epsilon) + P_{m-1}(\cos \epsilon) \bigr] =  2, \qquad 0 < x < \pi - \epsilon.
\end{align}
To prove this identity, we first use the Mehler's representation
(\ref{eq:mehler}) for Legendre polynomials $P_{m}(\cos \epsilon)$ and
$P_{m-1}(\cos \epsilon)$ to obtain
\begin{align}
S  =\frac{2}{\pi} \int^{\pi - \epsilon}_{0} 
\frac{
\sum^{\infty}_{m = 1} \bigl[ P_{m}(\cos x) + P_{m-1}(\cos x) \bigr] 
\left[ \cos (\left(m + \frac{1}{2}\right) t)-  \cos(\left(m - \frac{1}{2}\right) t)\right]  }{
\sqrt{(\cos(t)+\cos \epsilon)(\cos x- \cos(t))}
}, \qquad 0 < x < \pi.
\end{align}
We then use trigonometric identities and the series identity
(\ref{eq:Duffy}) to obtain the following integral representation
\begin{align}
S =  -\frac{2}{\pi} \int^{\pi - \epsilon}_{x} \! \! \! \mathrm{d}t \frac{2 \sin\left(\frac{t}{2}\right) 
\cos\left(\frac{t}{2}\right)}{\sqrt{(\cos(t)+\cos \epsilon)(\cos x- \cos(t))}}, \qquad 0 < x < \pi - \epsilon.
\end{align}
The consecutive change of variables $z = \cos(t)$ and $U =
\sqrt{(z+\cos \epsilon)(\cos x- z)}$ leads to
\begin{align}
S = \frac{2}{\pi} \int^{\cos x}_{-\cos x} \frac{\mathrm{d}z}{\sqrt{(z+\cos \epsilon)(\cos x- z)}} = 2,
\end{align}
which proves the identity (\ref{eq:fund}).  We will use this identity
in the following form:
\begin{align} \label{eq:fund2}
\sum^{\infty}_{m = 1} \bigl[ P_{m}(\cos x) + P_{m-1}(\cos x) \bigr] m \ \alpha_m =  1, \qquad 0 < x < \pi - \epsilon.
\end{align}
where $\alpha_m$ are given by Eq. (\ref{eq:ancarey}).

\subsubsection{Proof of the identity (\ref{eq:sumam0})}

We write $\alpha_m$ from Eq. (\ref{eq:ansinger}) and exchange the sum
and the integral to obtain:
\begin{align}
\sum^{\infty}_{m=1}  2 m \alpha^2_m = 
\frac{\sqrt{2}}{\pi} \int_{0}^{\pi - \epsilon} \! \! \! \mathrm{d}t \left(  \pd{}{t} \int_{0}^{t} \mathrm{d}x  
\frac{x \sin(x/2)}{\sqrt{\cos x - \cos t }} \right) \left( \sum^{\infty}_{m=1}  
\bigl[ P_m(\cos t) + P_{m-1}(\cos t) \bigr] m  \alpha_m \right). \label{eq:anan}
\end{align}
Using the identity (\ref{eq:fund}) in the right-hand side of
Eq. (\ref{eq:anan}) and the representation (\ref{eq:a0singer}) of
$\alpha_0$ leads to the result of Eq. (\ref{eq:sumam0}).

\subsubsection{An eigenvector of the matrix $M$} 

We show that the coefficients $\alpha_n$ (with $n \geq 1$) form an
eigenvector of the matrix $M$:
\begin{align} \label{eq:matrix_identity}
\sum^{\infty}_{m=1} M_{nm} \alpha_m= \alpha_n.
\end{align}
We express $M_{nm}$ through Eq. (\ref{eq:Mepsilon}) and exchange the
sum and the integral:
\begin{align}
\sum^{\infty}_{m=1} M_{nm} \alpha_m = 
 \frac{1}{2} \int^{1}_{-\cos\left( \epsilon\right)} \frac{1}{1+x} \left(P_n(x)+P_{n-1}(x)\right) \left( \sum^{\infty}_{m=1}  
\bigl[ P_m(\cos t) + P_{m-1}(\cos t) \bigr] m  \alpha_m \right)\mathrm{d}x,
\end{align}
from which the identity (\ref{eq:fund}) leads to the announced
identity (\ref{eq:matrix_identity}).

\section{Spatially averaged variances}
\label{sec:spatialaverage}

We denote by $\overline{\esp{\tau^n}}$ the spatial average of the
$n$-th moment of the exit time:
\begin{align} \label{def:averagednthmoment}
\overline{\esp{\tau^n}} \equiv \dfrac{1}{\pi} \int_{\vec{r}_0  \in \mathrm{\Omega}} \! \! \! \mathrm{d} \vec{r}_0 \ \esp{\tau^n_{\vec{r}_0}},
\end{align}
where $\mathrm{d} \vec{r}_0$ is the uniform measure over
$\mathrm{\Omega}$.
Let us now consider the random variable $\tau_{\mathrm{\Omega}}$,
defined in Sec. \ref{sec:recurrence} as the exit time of a particle
started at a random starting position $X$.  The $n$-th moment $\tau_{\mathrm{\Omega}}$
reads
\begin{align} \label{def:mfptaveraged}
\esp{\tau_{\mathrm{\Omega}}^n} \equiv \int_{\vec{r}_0 \in \mathrm{\Omega}} \! \! \! \mathrm{d}\mu(X = \vec{r}_0) \esp{\tau_{X}^n},
\end{align}
where $\mathrm{d} \mu(X =\vec{r}_0)$ is the probability density for
$X$ to be started at the position $\vec{r}_0$.  If $ \mathrm{d} \mu(X
=\vec{r}_0) =\mathrm{d} \vec{r}_0$ is the uniform probability
distribution, we can identify Eqs. (\ref{def:averagednthmoment}) and
(\ref{def:mfptaveraged}), $\esp{\tau_{\mathrm{\Omega}}^n} =
\overline{\esp{\tau^n}}$, and the variance of the random variable
$\tau_{\mathrm{\Omega}}$ is
\begin{align}
\Var \left[ \tau_{\mathrm{\Omega}} \right]  &\equiv \left( \int_{\vec{r}_0 \in \mathrm{\Omega}} \! \mathrm{d}\mu(X = \vec{r}_0) 
\esp{\tau_X^2} \right) - \left( \int_{\vec{r}_0 \in \mathrm{\Omega}} \! \mathrm{d}\mu(X= \vec{r}_0) \esp{\tau_X} \right)^2 
= \overline{\esp{\tau^2}} - \overline{\esp{\tau}}^2.
\end{align}
Note that $\Var \left[ \tau_{\mathrm{\Omega}} \right]$ differs from
the spatial average of the variance: $\Var \left[
\tau_{\mathrm{\Omega}} \right] \ne \overline{\Var \
\tau} = \overline{\esp{\tau^2}} - \overline{\esp{\tau}^2}$, because
$\overline{\esp{\tau}}^2 \ne \overline{\esp{\tau}^2}$.

\section{Convergence to an exponential distribution in the narrow-escape limit } \label{sec:conv}

\subsection{From the expression for the survival distribution} \label{sec:conv1}
We recall that the expressions for $(\alpha_n), n \geq 0,$ in the limit $\epsilon \ll 1$ are provided in Eqs. (\ref{eq:a0firstorder})--(\ref{eq:anfirstorder}). We denote by $\tilde{S}_{e}(t)$ the normalized single exponential
distribution whose mean is equal to the GMFPT defined in
Eq. (\ref{def:general_definition_average}).  The Laplace transform of
the distribution $\tilde{S}^{(t)}_{e}$ is
\begin{align} \label{eq:sedist}
S^{(p)}_{e} &= \dfrac{\overline{\esp{\tau}}}{1  + p \ \overline{\esp{\tau}}}.
\end{align}
We show that in the narrow-escape limit ($\epsilon \ll 1$), the
averaged exit time distribution $\overline{S^{(p)}} \approx
a^{(p)}_0/2$ converges to $S_{e}^{(p)}$, as expected from
Ref. \cite{Benichou:2010a}.  Due to the divergence of the coefficient
$\alpha_0$ from Eq. (\ref{eq:a0firstorder}), the asymptotic expansion
of Eq. (\ref{eq:a0exact}) at the first order in $\epsilon \ll 1$ reads
\begin{align} \label{eq:a0leading}
\frac{a_{0}^{(p)}}{2} &=  \alpha_0 \left[\pa_r f^{[1]}_{0}\right]_{\lvert r=1}  \left[ \sum^{\infty}_{k = 0} 
\left(p \left[\pa_r f^{[1]}_{0}\right]_{\lvert r=1} \alpha_0\right)^k  \right]+ \mathcal{O}(\epsilon) .
\end{align}
where $\left[\pa_r f^{[1]}_{0}\right]_{\lvert r=1}$ is the first-order
expansion in $p \ll 1$ of $\left[\pa_r f^{(p)}_{0}\right]_{\lvert
r=1}$.  At the leading order in $\epsilon$, the averaged survival
probability over $\mathrm{\Omega}$ is
\begin{align} \label{eq:spf}
\overline{S^{(p)}}  &= \frac{\alpha_0 \left[\pa_r f^{[1]}_{0}\right]_{\lvert r=1} }{1 + \alpha_0 \left[\pa_r f^{[1]}_{0}\right]_{\lvert r=1} } 
 + \mathcal{O}(\epsilon) .
\end{align}
Combining Eqs. (\ref{eq:spifunctionoff0}) and (\ref{eq:a0exact}), the
GMFPT at the leading order in $\epsilon \ll 1$ is
\begin{align}
\overline{\esp{\tau}} = \alpha_{0} \left[\pa_r f^{[1]}_{0}\right]_{\lvert r=1} + \mathcal{O}(\epsilon).
\end{align}
Combining Eqs. (\ref{eq:sedist}) and (\ref{eq:spf}) leads to
\begin{align}
\overline{S^{(p)}}  = \overline{S^{(p)}_{e}} + \mathcal{O}(\epsilon).
\end{align}
This shows the convergence in law of the FPT distribution to an
exponential distribution whose mean is the GMFPT as expected for the
narrow-espace limit \cite{Benichou:2010a}.

\subsection{From the expression for the moments} \label{sec:conv2}

Let us consider the random variable $\tau_{\mathrm{\Omega}}$, defined
in Sec. \ref{sec:recurrence} as the exit time of a particle started at
a random starting position.  We provide a positive answer to the
following question: does the distribution of $\tau_{\mathrm{\Omega}}$
converge to an exponential distribution in the limit $\epsilon \ll 1$,
even though the starting positions within the boundary layer
contribute to the statistics of $\tau_{\mathrm{\Omega}}$? Using the recurrence scheme
of Sec. \ref{sec:recurrence}, we verify that at the leading order in
$\epsilon \ll 1$ the moments of $\tau_{\mathrm{\Omega}}$ are
\begin{align}
\esp{\tau_{\mathrm{\Omega}}^n} = n! \left(\frac{\alpha_0}{2}\right)^n  + \mathcal{O}(\ln(\epsilon)^{n-1}),
\end{align}
which leads to $\esp{\tau_{\mathrm{\Omega}}^n} = n! \
\esp{\tau_{\mathrm{\Omega}}}~ (n \geq 1)$ at the leading order in
$\epsilon$.  The latter identity indicates that the FPT distribution
of $\tau_{\mathrm{\Omega}}$ converges to an exponential
distribution whose mean is the GMFPT defined by
Eq. (\ref{def:averaged1stmoment}), as expected from
Sec. \ref{sec:conv1}.  We emphasize that the relation
$\esp{\tau_{\mathrm{\Omega}}^n} = n! \ \esp{\tau_{\mathrm{\Omega}}}~
(n\geq 1)$, implies that the GMFPT characterizes the whole
distribution of the exit time, in contrast to the statement of
Eq. (18) from Ref. \cite{Mattos2012}.

\section{Computational aspects} \label{sec:app_montecarlo}

We summarize the numerical methods used to compute the FPT
distribution.  

(i) The Monte Carlo simulations rely on a sample of $2 \cdot 10^{6}$
of random walks.  This sample is obtained through $2$ hours of
computation on $200$ CPUs ($3.20$ GHz Intel Core{\texttrademark} i7).
The home-built C++ program uses an adaptive time step method so that
the time steps are given by a decreasing function with the distance to
the exit.

(ii) A finite element method realized in COMSOL Multiphysics v4.2
\cite{RogerW.Pryor2009} allowed to greatly reduce the computational time.  For
instance, the FPT probability density shown in Fig.\ref{fig:2}(a),(b)
required $5$ to $10$ minutes on a single CPU ($2,66$ GHz Intel Core
{\texttrademark} i5).  

(iii) The exact and approximate analytical solutions were computed in
MATLAB and using the numerical Laplace inversion package INVLAP
\cite{Valsa2011}.  The series were truncated at $N = 100$ terms and the
computational time is of the order of a few minutes on a single CPU
($2,66$ GHz Intel Core {\texttrademark} i5).

\bibliographystyle{unsrt}


\end{document}